\documentstyle[eqsecnum,aps,prb,multicol,epsf]{revtex}

\begin{document}
\newcommand{\fig}[2]{\epsfxsize=#1\epsfbox{#2}}
\def\asinh{{\rm asinh \ }}
\def\atan{{\rm atan \ }}

\title{Creep and depinning in disordered media}
\author{Pascal Chauve{$^{1}$}, Thierry Giamarchi{$^{1}$}
and Pierre Le Doussal{$^{2}$}}
\address{{$^1$} CNRS-Laboratoire de Physique des Solides,
Universit{\'e} de Paris-Sud, B{\^a}t. 510 , 91405 Orsay France}
\address{{$^2$} CNRS-Laboratoire de Physique Th{\'e}orique de
l'Ecole Normale Sup{\'e}rieure, 24 rue Lhomond, 75231 Cedex 05, Paris France.}
\date{\today}
\maketitle

\begin{abstract}
Elastic systems driven in a disordered medium exhibit a depinning
transition at zero temperature and a creep regime at finite
temperature and slow drive $f$. We derive functional renormalization group
equations which allow to describe in details the properties of
the slowly moving states in both cases. Since they hold at finite velocity $v$,
they allow to remedy some shortcomings of the previous approaches to
zero temperature depinning. In particular, they enable us to derive the
depinning law directly from the equation of motion, with no artificial
prescription or additional physical assumptions, such as a scaling
relation among the exponents. Our approach provides a
controlled framework to establish under which conditions the depinning regime
is universal. It explicitly demonstrates that the random
potential seen by a moving extended system evolves at large
scale to a random field and yields a self-contained picture for
the size of the avalanches associated with the deterministic motion.
At finite temperature $T>0$ we find that the effective barriers
grow with lenghtscale as the energy differences between neighboring
metastable states, and demonstrate the resulting activated creep law
$v\sim \exp \left(-C\, f^{-\mu}/T \right)$ 
where the exponent $\mu$ is obtained in a
$\epsilon=4-D$ expansion ($D$ is the internal dimension of the
interface). Our approach also provides quantitatively a new scenario
for creep motion as it allows to identify several intermediate lengthscales.
In particular, we unveil a novel ``depinning-like'' regime at scales
larger than the activation scale, with avalanches spreading from the
thermal nucleus scale up to the much larger correlation length
$R_{V}$. We predict that $R_{V}\sim T^{-\sigma }f^{-\lambda }$ 
diverges at small drive and temperature with exponents $\sigma ,\lambda $ 
that we determine.
\end{abstract}

\begin{multicols}{2}

\section{Introduction}
\label{introduction}

Understanding the statics and dynamics of elastic systems 
in a random environment is a longstanding problem with important 
applications for a host of experimental systems. Such problems can be split 
into two broad categories: (i) propagating interfaces such as magnetic domain 
walls\cite{lemerle_domainwall_creep}, fluid invasion in porous media 
\cite{wilkinson_invasion} or epitaxial growth \cite{barabasi_book}; 
(ii) periodic systems 
such as vortex lattices\cite{blatter_vortex_review}, charge density
waves\cite{gruner_revue_cdw}, or Wigner crystals of
electrons\cite{andrei_wigner_2d}.
In all these systems the basic physical ingredients are identical:
the elastic forces tend to keep the 
structure ordered (flat for an interface and periodic for lattices),
whereas the impurities locally promote the wandering. From the
competition between disorder and elasticity emerges a complicated
energy landscape with many metastable states. This results in glassy
properties such as hysteresis and history dependence of the static
configuration. In the dynamics, one expects of course this 
competition to have important consequences on the response of the
system to an externally applied force. 

To study the statics, the standard tools of
statistical mechanics could be applied, leading to a good understanding of 
the physical properties. Scaling arguments and simplified models showed 
that even in the limit of weak disorder, 
the equilibrium large scale properties of disordered elastic systems
are governed by the presence of impurities. In particular, below four
(internal) dimensions, displacements grow unboundedly\cite{larkin_70}
with the distance, resulting 
in rough interfaces and loss of strict translational order 
in periodic structures\cite{blatter_vortex_review}. To go beyond
simple scaling arguments and obtain a more detailed description
of the system is rather difficult and at present only main two methods,
each with its own shortcomings, have been developped. The first one
is to perform a perturbative renormalization group calculation on the disorder,
and is valid in $4-\epsilon$ dimensions to first order in $\epsilon$.
In this functional renormalization group (FRG) approach 
\cite{fisher_functional_rg,balents_frg_largen}, 
the whole correlation function of the disorder is renormalized.
The occurence of glassiness is signalled by a non-analyticity
appearing at a finite lengtscale during the flow, specifically a
cusp in the force correlator. This yields non trivial predictions
for the roughness exponents of interfaces\cite{fisher_functional_rg}.
Another approach relies
on the replica method to study either the mean field limit
(i.e large number of components) or to perform a gaussian variational
approximation of the physical model. Using this variational approach
both for manifolds \cite{mezard_variational_replica}
and for periodic 
systems\cite{bragg_glass_global,korshunov_variational_short}, correlation 
functions and thermodynamic properties could be obtained. It confirms
the existence of glassy properties, with energy fluctuations 
growing as $L^\theta$ where $\theta$ is a positive exponent. To obtain
the glass phase in this method, one must break the replica symmetry. 
At a qualitative level, this is in good agreement
with the physical intuition of such systems as being composed of
many low lying metastable states separated by high barriers.
As was clearly shown in the case of periodic manifolds, the correlation
functions can be obtained by both the FRG and variational approach
and are found to be in very reasonable agreement, bridging the 
gap between the two methods \cite{bragg_glass_global,balents_rsb_frg}.
Taken together, these two approaches thus provide a very coherent picture for the 
statics\cite{giamarchi_book_young,balents_rsb_frg}.
In particular they allow to understand that although disorder 
leads to glassy features in both the manifold and the periodic systems, 
these two types of problems belong to quite different universality classes 
in other respects, such as the large distance
behaviour of the correlations \cite{giamarchi_book_young}.

\begin{figure}[htb]
\centerline{\fig{6cm}{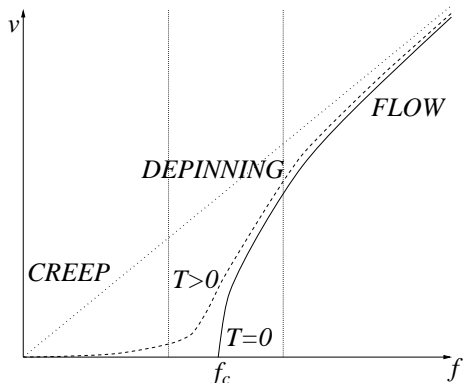}}
\caption{{\narrowtext Typical force--velocity characteristics,
exhibiting pinning at $T=0$ with a threshold force $f_{c}$ and creep
at $T>0$, $f<f_{c}$. At large drive, the system flows as if there were
no disorder.}}
\label{vfinsp}
\end{figure}

These properties have drastic consequences for the dynamics of driven systems
in the case, important in practice, where an elastic description holds
(i.e when plastic deformations can be neglected). Determining 
the response to an externally applied force is not only an interesting 
theoretical question, but also one of the most important experimental 
issues. Indeed in most systems the velocity $v$ versus force $f$ 
characteristics is directly measurable and is simply linked to
the transport properties
(voltage-current for vortices, current-voltage for CDW and Wigner crystals, 
velocity-applied magnetic field for magnetic domain walls).
In the presence of disorder it is natural to expect that,
at zero temperature, the system remains pinned and
only polarizes under the action of a small applied force, i.e. moves until it
locks on a local minimum of the tilted energy landscape. At larger drive, the
system follows the force $f$ and acquires a non-zero asymptotic
velocity $v$. In the simplest cases, the effect of disorder at large velocity
is washed out and one recovers the viscous flow, as in the pure case. In the
thermodynamic limit, it is believed that there exists a threshold
force $f_{c}$ separating both states, and that a dynamical
transition occurs at $f_{c}$ called {\it depinning}, where the
velocity is continuously switched on, like an order parameter of a
second order transition in an equilibrium
system\cite{fisher_depinning_meanfield}, leading to a $v$--$f$
characteristics such as the one shown in Figure~\ref{vfinsp}. 
An estimate of $f_c$ can be obtained via scaling arguments
\cite{larkin_ovchinnikov_pinning} or with a criterion for the
breakdown of the large velocity expansion\cite{schmidt_hauger,larkin_largev}.
Beyond $f_c$, if one describes the depinning as a conventional
dynamical critical
phenomenon, the important quantities to determine are of course 
the depinning exponent $\beta$ giving the velocity 
$v \sim (f-f_c)^\beta$ and the dynamical exponent $z$
which relates space and time as $t\sim r^z$.

An even more challenging question, and experimentally at least as relevant,
is the response at finite temperature $T>0$. In the most naive description,
the system can now overcome barriers via thermal activation, leading to a 
thermally assisted flow \cite{anderson_kim} 
and a linear response at small force of the 
form $v \sim e^{-\Delta /T} f$, where $\Delta$ is some typical barrier.
It was realized 
\cite{nattermann_rfield_rbond,ioffe_creep,nattermann_creep_domainwall,%
feigelman_collective} that
because of the glassy nature of the static system, the motion is
actually dominated by barriers which {\it diverge} as the drive $f$
goes to zero, and thus the flow formula with finite barriers
is incorrect. Well below the threshold critical force, the barriers are
very high and thus the motion, usually called ``creep'' is extremely slow.
Scaling arguments, relying on strong assumptions such
as the scaling of energy barriers and the use of statics properties to describe 
an out of equilibrium system, were used to infer the small $f$ response.
This led to a non linear response, characteristic of the creep
regime, of the form $v\sim \exp (- C\, f^{-\mu}/T)$
where $\mu=(D-2+2\zeta_{\rm eq})/(2-\zeta_{\rm eq})$ and $\zeta_{\rm eq}$ is 
the roughening exponent for the static $D$-dimensional system.

Obtaining a detailed experimental confirmation of this behaviour
is a non trivial feat, in reasons of the range in velocity required.
Although in vortex systems these highly non linear flux creep behaviours
have been measured ubiquitously,
it is rather difficult to obtain clean
determination of the exponents, given the many regimes of lengthscales
which characterize type II superconductors
\cite{blatter_vortex_review}. In some recent 
measurement, some agreement with the creep law in the Bragg
glass regime was obtained\cite{fuchs}.
Probably the most conclusive evidence for the above law was
obtained, not in vortex systems, but for magnetic interfaces. 
Quite recently Lemerle {\it et al.}\cite{lemerle_domainwall_creep} 
successfully fitted the force-velocity
characteristics of a magnetic domain wall driven on a random
substrate by a stretched exponential form $v\sim \exp -f^{-0.25}$ over
{\it eleven} 
decades in velocity. This provided evidence not only of the stretched 
exponential behavior, but of the validity of the exponent as well. 

Given the phenomelogical aspect of these predictions and the uncontrolled
nature of the assumptions made, both for the 
creep and for the depinning, it is important to derive this
behavior in a systematic way from the equation
of motion. Less tools are available than for the statics,
and averages over disorder should be made using dynamical methods.
Fortunately, it is still possible to use a functional renormalization
group (FRG) approach for the dynamical problem. Such an 
approach has been used at $T=0$ to study depinning 
\cite{narayan_fisher_cdw,nattermann_stepanow_depinning}.
It allowed for a calculation of the depinning exponents, in $D=4-\epsilon$.
However this approach is still rather unsatisfactory. The 
FRG flow used in \cite{narayan_fisher_cdw,nattermann_stepanow_depinning}
is essentially the static one, the finite velocity being only invoked
to remove - by hand - some ambiguities and to cutoff the flow, with no 
real controlled way to show that this is the correct procedure.
Furthermore in these approaches
it is also necessary to assume, instead of deriving them from the FRG,
some scaling relations in order to obtain the exponents.
Another rather problematic point is that, with no additional
input, the method of \cite{nattermann_stepanow_depinning}
would yield three universality classes for the depinning: two
universality classes
depending on the nature of the disorder (random bond versus random field)
for manifolds and one for periodic systems, while numerics and physical
arguments \cite{narayan_fisher_cdw} suggested 
that only two (random field and periodic) universality classes could 
exists. In addition, since this is also intrinsically a $T=0$ (and $v=0$)
approach, it can not be used to tackle the creep behavior.

We propose here a single theory for describing all the regimes of a
moving elastic system, including depinning and the non-zero temperature
regimes. Our FRG equations contain from the start the finite velocity
and finite temperature. They thus allow to address questions which
are beyond the reach of either approximate scaling theories,
or $v=0$ FRG flow. For the depinning we are able to
determine the conditions required for the existence of
a universal depinning behavior, as well as computing the depinning 
exponents (and estimating $f_c$). We show in particular that only 
two universality classes exist (out of the three) for the depinning
since we explicitly find that random bond systems flow to the random
field universality class. We can also extract from our equations the 
characteristic lengthscales of the depinning. The main advantage of 
our approach is of course to address the finite $T$ small $v$ regime 
as well. The method allows to {\it derive} the creep formula directly
and thus allows to confirm the assumptions made on the scaling
of the energy barriers. In addition we show that the creep is followed
by a depinning-like regime and determine its characteristic lengthscales.
A short account of some of these results 
was presented in Ref.~\onlinecite{chauve_creep_short}.

The paper is organized as follows: in Section~\ref{elasticitydisorder}
we present the equation of motion and the types of disorder studied here.
Section~\ref{preliminary} is devoted to a brief review of scaling
arguments and a summary of useful results from perturbation theory,
presented in Appendix~\ref{app:pt}.
Section~\ref{fieldtheory} contains the
field theoretical formulation of the problem and the associated 
renormalization group flow equations, derived in 
Appendix~\ref{app:derivation}. The static case is studied in
Subsection~\ref{statics}, focusing on the appearance of the cusp. The
effect of the temperature is studied in details in Appendices
\ref{app:temperature} and \ref{app:ftfp}. In the next sections, we study
the depinning (\ref{depinning}) and creep (\ref{creep}) regimes. Both
sections contain the outline of the derivation and a physical
discussion. Appendix~\ref{app:v} is devoted to the effect of a small
velocity on the FRG. We conclude in Section~\ref{conclusion}, referring to an
extension of our work proposed in Appendix~\ref{app:n>1}.
In Appendix~\ref{app:notations} we fix the notations used
throughout the paper.

\section{Elasticity and disorder}
\label{elasticitydisorder}

Elastic systems are extended objects which ``prefer'' to be flat or
well ordered. We are dealing with two different types of elastic
systems which however can be treated in the same way. On the one
hand, {\it interfaces}, i.e. surfaces with a stiffness that makes local
distortions energetically expensive, on the other hand, {\it lattices}
with elastic displacements allowed about a regularly ordered configuration.

\begin{figure}[h]
\centerline{\fig{6cm}{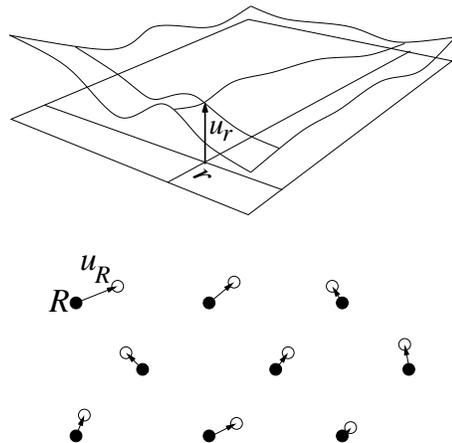}}
\caption{{\narrowtext Top: an interface with height field
$u_{r}$ above $r$. We denote by $r$ the (internal) coordinates along the
interface and by $z$ the height coordinates.
Bottom: a lattice with reference positions $R$ and displacements $u_{R}$
from $R$.}}
\label{interface}
\end{figure}

The first type is the easiest to visualize. 
The interface is assumed to have no overhangs and is thus described by a
height function $u_{r}$ defined at each point $r$ (see
Figure~\ref{interface}). Its
energy is proportional to its area $\int_{r} \sqrt{1+|\nabla u|^{2}}$
and in the elastic limit $|\nabla u|\ll 1$, reduces to
\begin{eqnarray} \label{elastic}
H_{\rm el}[u]=\int_{r} \frac{c}{2}|\nabla u|^{2}
\end{eqnarray}
relative to the flat $u_{r} = 0$ configuration (notations are defined
in the Appendix~\ref{app:notations}). We denote
by $c$ the stiffness, or elastic constant.
Periodic structures, such as flux line
lattices or charge density waves (CDW), can be described by the same type
of elastic Hamiltonian. For each point (or line) in the elastic
periodic system one can introduce a (vector) displacement field
$u_{R}$ that gives the shift from the reference position $R$ (see
Figure~\ref{interface}).
The elastic energy for small displacements
is given by a quadratic form in the differences
$u_{R}-u_{R'}$ between neighboring points and thus can be written
as (\ref{elastic}) in a continuum
description ($r$ being a generic point in space). When $u$ has more
than one component, $c$ should be understood as a tensor (see
Appendix~\ref{app:n>1}).

To take the quenched disorder into account in such systems it is necessary
to express the energy of the above elastic structure in the presence
of impurities. The coupling to a substrate or to local fields is
easily written for interface models and is more subtle for
lattices. Quite generally the coupling to disorder leads to an energy:
\begin{equation} \label{disorgeneral}
H_{\rm dis} = \int_{r}V(r,u_{r})
\end{equation}
which gives rise to a pinning force $F (r,u)=-\partial_{u}V
(r,u)$ acting on the displacement $u_{r}$. Depending on the
microscopic origin of the disorder term $V$, the coupling 
(\ref{disorgeneral}) leads to quite different physics. 

In the case of interfaces (\ref{disorgeneral}) originates from
\begin{eqnarray}  
H_{\rm dis} &=& \int_{r,z} V(r,z) \rho(r,z)\label{hdisinterface}\\
\rho(r,z) &=& \int_{\kappa_z} e^{i \kappa_z \cdot 
(z - u_r)}=\delta (z-u_{r})\label{rhointerface}
\end{eqnarray}
in terms of the density $\rho (r,z)$.
One then usually distinguishes two cases: either ``random bond'' (RB)
when $V (r,z)$ is short range (random exchange for magnetic domain
walls), or ``random field'' (RF) as discussed below, where $V (r,z)$ has
long range correlations.

In the case of periodic structures, the density $\rho (r)$ can be
expressed using the set of vectors $\kappa $ of the reciprocal lattice
and (\ref{disorgeneral}) originates from
\begin{eqnarray} 
H_{\rm dis} &=& \int_{r} W(r) \rho(r)\label{hdislattice}\\
\rho(r) &\simeq & \rho_0\sum_{\kappa } e^{i \kappa \cdot(r
- u_r)}\label{rholattice}
\end{eqnarray}
where $\rho_0$ the average density\cite{bragg_glass_global}.
The potential $W$ is random, of short range $r_{f}$ (e.g. 
point impurities for a
vortex lattice or a CDW). We call this case ``random periodic'' (RP).

In both cases, using
(\ref{hdisinterface},\ref{rhointerface},\ref{hdislattice},\ref{rholattice})
and (\ref{disorgeneral}) one obtains for the correlations 
of $V$ in (\ref{disorgeneral}) 
\begin{equation}\label{r}
\overline{(V (r,u)-V (r',u'))^{2}}=-2\delta _{rr'}{\sf R} (u-u')
\end{equation}
where ${\sf R} (u)$ is a periodic function with the periodicity $a$
of the lattice in the periodic (RP) case\cite{arf}. 
The $\delta$ function is cutoff
at the microscopic scale $r_{f}$. 

\begin{figure}[htb]
\centerline{\fig{6cm}{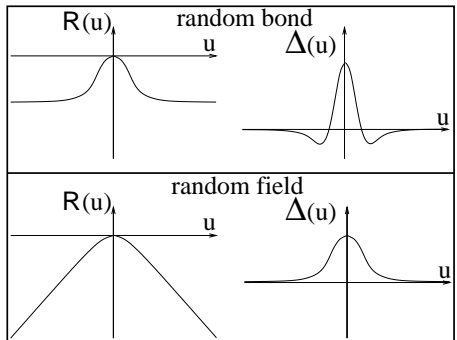}}
\caption{{\narrowtext up: RB case, down: RF case. Right: correlator of
the potential, left: correlator of the force.}}
\label{rbrfsketch}
\end{figure}

For an interface, ${\sf R} (u)$ has the shape shown in
Figure~\ref{rbrfsketch}. In that case, 
the width $r_f$ of ${\sf R} (u)$ is typically given by the width of the
interface or the size of impurities.
The force resulting from such a random bond disorder 
has correlations\cite{asymmetry}
\begin{eqnarray}\label{delta}
\overline{F (r,u)F (r',u')}=\delta _{rr'} \Delta(u-u')
\end{eqnarray}
as shown in Figure~\ref{rbrfsketch} where
\begin{equation} \label{deltar}
\Delta (u)=-{\sf R}''(u)
\end{equation}
The signature of such a RB
disorder for the interface is that $\int \Delta = 0$ since $R' (u)$
decreases to zero at infinity.

Another type of disorder occurs in the case of interfaces separating
two phases, like e.g. a domain wall in a disordered magnet. A random
field couples
differently to the two phases on the right and left of the interface, thus
the energy resulting from the coupling to disorder involves an integral in
the bulk of the system and not just {\it at} the interface position.
The correlation of the force can still be expressed by (\ref{delta})
and $\Delta$ still decreases to zero above a scale $r_f$ as shown 
on Figure~\ref{rbrfsketch}.
Contrarily to the RB case, $\int
\Delta$ does not vanish.  For a single
component displacement field $u$, the RF, of correlator (\ref{delta}),
is still  
{\it formally} the derivative of a potential $V(r,u)=-\int ^{u}du'\,F
(r,u')$. The correlations of this fictious
potential are of the form (\ref{r}) with 
${\sf R} (u)=-\int_{0}^{u}du'\,\int_{0}^{u'}du''\,\Delta (u'')$,
and one has ${\sf R} (u)\simeq -\frac{1}{2}|u|\int \Delta $ for $|u|\gg
r_{f}$ which can be visualized as a random walk (where $u$ plays the
role of ``time'' and the {\it random field strength} $\int \Delta
=-2{\sf R}' (\infty )$ is the ``diffusion constant'').
Contrarily to the RB for which ${\sf R}(u)$ is short range, 
${\sf R}(u)$ for the RF grows at large $u$ as shown on Figure~\ref{rbrfsketch}.

In this paper we study the overdamped driven motion of such elastic
systems which obey 
\begin{equation} \label{notcomoving}
\eta \partial_{t}u_{rt} = c\nabla ^{2}u_{rt}+F (r,u_{rt}) + \zeta _{rt}+ f
\end{equation}
where $\eta $ is a friction,
$f$ is the external driving force density and $\zeta _{rt}$ a Langevin
noise. The correlation $\langle \zeta_{rt} \zeta_{r't'} \rangle =2\eta
T\delta _{rr'}\delta _{tt'}$ defines as usual a temperature $T$ for this
out of equilibrium system. The long time behavior of (\ref{notcomoving})
at zero drive 
$f=0$, reduces to the thermodynamics at temperature $T$. In 
(\ref{notcomoving}) the bare\cite{bare} 
pinning force $F (r,u)$ is gaussian with
zero average and correlator given by (\ref{delta}). We will consider 
three universality classes for $\Delta $ corresponding
to an interface in a random potential (RB), in a random
field (RF) or a periodic system in a random potential (RP). Physical
realizations of such disorders would be respectively a random anisotropy
for a magnetic domain wall \cite{lemerle_domainwall_creep}, the random
field Ising systems \cite{nattermann_book_young} and vortex lattices or CDW
\cite{bragg_glass_global,narayan_fisher_cdw}.

It is also useful to rewrite (\ref{notcomoving}) in the co-moving
frame at average velocity 
$v=\overline{\langle \partial_{t}u_{rt}\rangle }$. In the remainder of
this paper, we switch to $u_{rt}\rightarrow u_{rt}+vt$ and thus study
the following equation of motion
\begin{eqnarray}\label{comoving}
\left\{\begin{array}{rcl}
\overline{\langle \partial_{t}u_{rt}\rangle }&=&0\\
(\eta \partial_{t}-c\nabla ^{2})u_{rt}&=&F (r,vt+u_{rt})+
\zeta _{rt}+\tilde{f} \end{array} \right.
\end{eqnarray}
where $\tilde{f}=f-\eta v$ is the average pinning force and $r$
belongs to a $D$-dimensional internal space. From now on
we specialize to an unidimensional displacement field $u_{r}$ as
would be the case for an interface model or a single $Q$ CDW. This
simpler case already captures the main physics at small
velocity, investigated here. 
Extensions to many-component systems will be briefly discussed.

Before giving a quantitative treatment using renormalization group,
let us review the qualitative arguments which have been given
previously to describe the physics originating from (\ref{notcomoving}).

\section{Preliminary arguments}
\label{preliminary}

\subsubsection{Statics}
\label{scstatics}

In the absence of drive, (\ref{notcomoving}) is equivalent to 
the equilibrium problem at temperature $T$.
The state of the system results from
the competition between elasticity, pinning and thermal fluctuations. 
The physics of such
problems can be investigated by a host of methods
\cite{blatter_vortex_review,bragg_glass_global,%
larkin_ovchinnikov_pinning,larkin_70,%
fisher_functional_rg}
and here we only recall the salient points. Temperature does not play
an important role as will become clear and we begin with the $T=0$ case.

A subsystem of size $R$, with displacement
$w(R)=\sqrt{\overline{\left(u_{R}-u_{0} \right)^{2}}}$, 
is submitted to a typical elastic
force density $f_{\rm el}=c w (R) /R^{2}$ and to a typical pinning force
density $f_{\rm pin}=\sqrt{\Delta (0)/R^{D}}$. Balancing these
quantities, one obtains that elasticity wins at large scales for
$D>4$, resulting in a flat interface with a priori bounded
displacements. In $D<4$, systems of
size $R$ smaller than the Larkin length 
\begin{eqnarray}\label{larkinlength}
R_{c}=\left(\frac{c^2 r_f^2}{\Delta (0)}
\right)^{\frac{1}{4-D}}
\end{eqnarray}
wander as predicted by the Larkin model\cite{larkin_70}: 
\begin{eqnarray}\label{larkinw}
w(R) \sim r_{f}\left(\frac{R}{R_{c}} \right)^{\frac{4-D}{2}}
\end{eqnarray}
At larger 
scales $R>R_{c}$, the system wanders 
further than the correlation length $r_f$ of the disorder. This simple
picture breaks down and the system 
can be viewed as made of Larkin
domains of size $R_{c}$, which are independently pinned. First order
perturbation theory confirms this picture below the Larkin length. 
The static equilibrium (equal-time) correlation function
at $T=0$ is (see Appendix~\ref{app:pt})
\begin{eqnarray}\label{unsurq4}
\overline{u_{-q,t}u_{q,t}}=\frac{\Delta (0)}{(cq^{2})^{2}}
\end{eqnarray}
The wandering computed from (\ref{unsurq4}) 
\begin{eqnarray}
\frac{1}{2}\overline{\left(u_{r,t}-u_{0,t} \right)^{2}}
\sim \frac{\Delta (0)}{c^{2}}
S_{4}r^{\epsilon }/\epsilon
\end{eqnarray}
for $D=4-\epsilon $ gives back (\ref{larkinw}), and we recover the
scaling expression 
(\ref{larkinlength}) by equating the
wandering to $r_{f}^{2}$:
\begin{eqnarray}\label{larkinlength2}
R_{c}=
\left(\epsilon \frac{c^{2}r_{f}^{2}}{S_{4}\Delta (0)} \right)^{1/\epsilon }
\end{eqnarray}
We used that $\int_{q}\frac{1-\cos q.r}{q^{4}}=A_{D}r^{4-D}$
for $2<D<4$ with
$A_{D=4-\epsilon }=-\pi ^{\epsilon /2-2} \Gamma[-\epsilon /2]/16
\sim S_{4}/\epsilon $ when $\epsilon \rightarrow 0^{+}$.

The remarkable feature is that $T=0$ 
perturbation theory (either using replicas or
equilibrium dynamics) gives that (\ref{unsurq4}) is exact {\it to all
orders}
\cite{efetov_larkin_replicas,chauve_frg_futur} in $\Delta$, 
and is identical to
the correlation in the Larkin
model\cite{larkin_70}. Indeed, the naive perturbation series
organizes as if the pinning
energy were simply expanded in $u$ (thus the pinning force is
independent of $u$ with $\overline{F (r)F (r')}=\Delta (0)\delta
(r-r')$), resulting in a gaussian model.

In fact, due to the occurence 
of multiple minima beyond
$R_{c}$, this perturbative result is incorrect\cite{villain_semeria} 
at large scale. It
can be shown, for example on discrete systems, that if a 
configuration $u^{\rm GS}_{r}$ which minimizes
$H[u]=\int_{r}\left[\frac{c}{2} (\nabla u_{r})^{2}+V
(r,u_{r})\right]$ is defined on a volume larger than $R_{c}$, then
the Hessian $\frac{\delta^{2}H}{\delta u_{r}\delta u_{r'}}[u^{\rm
GS}]$ becomes singular \cite{tanguy_phd}.
Such instability appears clearly in a functional renormalization group (FRG)
treatment of the problem\cite{fisher_functional_rg} which proves that
$\Delta$ becomes nonanalytic beyond the length $R_c$, as will be
discussed below. It can also be seen within variational or
mean-field treatments using replicas \cite{mezard_variational_replica}
that replica symmetry breaking (RSB) is
necessary to describe the physics beyond the Larkin length $R_c$. 
Using either replicas with RSB or the FRG it is possible to
describe the physics at all scales and to obtain the correct roughness
exponent $\zeta_{\rm eq}$ defined by
\begin{equation}
w (R)\sim r_f \left(\frac{R}{R_{c}} \right)^{\zeta_{\rm eq}}
\end{equation}
where the value of $\zeta_{\rm eq} $ depends on the statics universality
class\cite{blatter_vortex_review}. 
Since disorder induces unbounded displacements, 
the system is rough and the temperature is always 
formally irrelevant in $D> 2$. It is described by a $T=0$ fixed
point, characteristic of a glass phase.

\subsubsection{Depinning}
\label{scdepinning}

An elastic system does not necessarily move under the action of a
driving force. The disorder 
leads to the existence of a threshold force $f_c$ at $T=0$
as shown on Figure~\ref{vfinsp}. A simple dimensional estimate of
$f_c$ can be obtained\cite{larkin_ovchinnikov_pinning} by computing
the sum of the
independent pinning forces acting on the
Larkin domains 
$(R/R_{c})^{D} \sqrt{\Delta (0) R_{c}^{D}}$ and balancing it with the
driving force acting on the same volume
$R^{D}f$. This gives
\begin{equation} \label{fclarkin}
f_{c}\sim \frac{c\, r_{f}}{R_{c}^{2}}
\end{equation}
Another estimate of $f_c$ comes from the large velocity expansion
\cite{larkin_largev,schmidt_hauger} of the equation of motion
(\ref{comoving}) (from
the criterion $\tilde{f}\simeq \eta v$). It coincides with 
(\ref{fclarkin}).

For $f\gtrsim f_c$ the system moves with a small velocity, and it has
been proposed\cite{fisher_depinning_meanfield} that depinning can be 
described in the framework of standard critical phenomena, 
with the velocity as an order parameter. This leads to the assumption
of two independent critical 
exponents\cite{nattermann_stepanow_depinning,narayan_fisher_cdw,%
narayan_fisher_depinning} 
$\zeta$ and $z$, 
defined through the correlation function in the comoving frame 
(in the stationary state for $f\rightarrow f_{c}^{+}$)
\begin{equation}
\overline{(u_{r,t}-u_{0,0})^{2}}=r^{2\zeta}{\cal C}
(t/r^{z})
\end{equation}
${\cal C} (x)\rightarrow {\rm cst}$ for $x\rightarrow 0$
and ${\cal C} (x)\sim x^{2 \zeta/z}$ for $x\rightarrow \infty $. The
dynamical roughening exponent $\zeta$ close to the threshold a priori
differs from its equilibrium value $\zeta_{\rm eq}$. 
Several related exponents can be also
introduced such as: (i) the depinning exponent $\beta$;
(ii) the correlation length exponent $\nu$ describing the divergence of the
length $\xi$ defined from the equal-time velocity-velocity correlation
function. They satisfy
\begin{eqnarray}
v &\sim& (f-f_c)^\beta \\
\xi&\sim& (f-f_{c})^{-\nu}
\end{eqnarray}
Numerically\cite{tang_leschhorn_simus} the motion of the system
looks like a deterministic succession of
avalanches of size $\xi$ with characteristic time $\tau\sim
(f-f_{c})^{-z \nu}$. 
From the argument $u\sim v\tau $ and the statistical tilt
symmetry\cite{nattermann_stepanow_depinning,narayan_fisher_cdw} 
(see below), the exponents $\beta $ and $\nu $ are
usually determined from $\zeta ,z$ by the scaling relations
\begin{equation}
\nu = \frac1{2-\zeta} =\frac{\beta}{(z-\zeta)}
\end{equation}

To obtain these exponents analytically, one needs to perform an FRG
analysis of the equation of motion. This will be discussed in more
details in Section~\ref{depinning}.

\subsubsection{Creep}
\label{sccreep}

At finite temperature $T>0$, motion occurs at any drive.
For low temperatures and very small drive $f\ll f_c$ one
expects the motion to be very slow, and thus, although it
is a dynamical problem, a qualitative understanding 
can be obtained by considering
thermal activation over barriers determined from {\it statics}
arguments. An original estimate\cite{anderson_kim} of such barriers
led to linear, albeit activated, response. However the effects linked
to the glassy nature of the problem were understood at a qualitative
level\cite{nattermann_rfield_rbond,ioffe_creep,nattermann_creep_domainwall,%
feigelman_collective} using scaling arguments. 

\begin{figure}[htb]
\centerline{\fig{6cm}{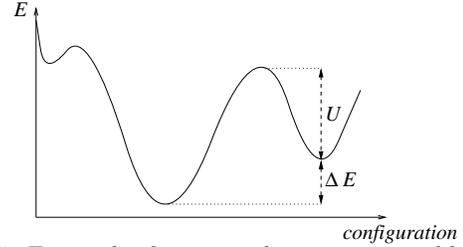}}
\caption{{\narrowtext Energy landscape, with many metastable states in
the valleys, differing by $\Delta E$, and barriers $U$ between them.}}
\label{landscape}
\end{figure}

The argument proceed as follows:
systems larger than $R_{c}$ have a (static) roughness $w (R)\sim
r_{f}\left(\frac{R}{R_{c}} \right)^{\zeta _{\rm eq}}$ and hence the energy has
typical fluctuations of order
\begin{equation}
E (R)\sim U_{c}\left(\frac{R}{R_{c}} \right)^{D-2+2\zeta _{\rm eq}}
\end{equation}
with $U_{c}=cR_{c}^{D-2}r_{f}^{2}$ the energy scale 
of a Larkin domain. Assuming
that the energy landscape is characterized by a {\it unique} energy
scale, and thus that the energy differences between neighbouring
metastable states is the same as the energy barrier separating them
as schematically shown in Figure~\ref{landscape}, one obtains that
the barriers height scale with an exponent $D-2+2 \zeta _{\rm eq}$.
Since the motion is very slow, it is usually argued that 
the effect of the drive is just to tilt the
energy landscape, and the effective barrier becomes
\begin{equation}\label{barriertilt}
U_{c}\left(\frac{R}{R_{c}}
\right)^{D-2+2\zeta _{\rm eq}}-fR^{D}r_{f}\left(\frac{R}{R_{c}}
\right)^{\zeta _{\rm eq}}
\end{equation}
The maximum of (\ref{barriertilt}), obtained at $R_{\rm opt}\sim
R_{c}\left(\frac{f}{f_{c}} \right)^{-\frac{1}{2-\zeta _{\rm eq}}}$, gives via
Arrhenius law the largest time spent in the valley
by the thermally activated system and thus yields the velocity
\begin{eqnarray}
v\sim \exp \left[-\frac{U_{c}}{T}\left(\frac{f}{f}_{c}
\right)^{-\mu}\right] \qquad
\mu=\frac{D-2+ 2 \zeta _{\rm eq}}{2-\zeta _{\rm eq}}
\end{eqnarray}
known as the {\it creep} motion, characterized by the stretched
exponential with exponent $\mu$. Note that the effective barrier given
by the above formula vanishes at a scale $R_{0}\sim
R_{c}\left(\frac{f}{f_{c}} \right)^{-\frac{1}{2-\zeta _{\rm eq}}} $ which
diverges as fast as $R_{\rm opt}$, the typical size of a thermally
activated excitation.
\begin{figure}[htb]
\centerline{\fig{6cm}{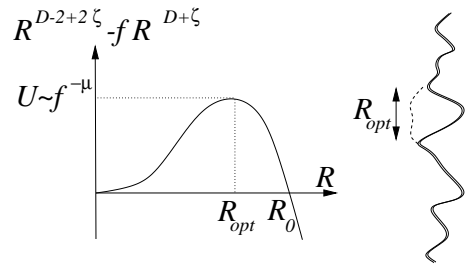}}
\caption{{\narrowtext Effective barrier and motion by nucleation.}}
\label{nucleation}
\end{figure}

This elegant scaling argument leading to the creep formula relies however
on strong assumptions and does not yield any information on the
detailed behavior, in particular on what happens after the thermal
jumps. The fact that {\it static} barriers and valleys scale with
the same exponent is already a non-trivial hypothesis about the structure of
the infinite-dimensional energy landscape. Refined 
simulations\cite{drossel_barrier,mikheev_barrier_2d,drossel_barrier_3d}
of a directed polymer $D=1$ in $N=1$ 
and $N=2$ are consistent with the ``equal 
scaling assumption'' for this particular case, but a general proof is still
lacking. The second, and more delicate hypothesis is the validity of
the Arrhenius description: (i) the system being out of equilibrium, it
is not clear that {\it dynamical} barriers can be determined
purely from the statics; (ii) one assumes that the
motion is dominated by a {\it typical} barrier.
These assumptions can turn incorrect for some specific problems.
For example, in the case of a point
moving in a one-dimensional random potential, 
the $v$-$f$ characteristics at low drive is {\it
not} \cite{ledoussal_anomalous_diffusion} of Arrhenius type.
Although this $0+1$ case is peculiar since the particle has no freedom
to pass aside impurities (it is dominated at $T=0$ by the highest
slope of the potential and at finite $T$ by the rare highest barriers),
one should also address the question of the distribution of barriers in higher
dimensions.

\section{Dynamical action and renormalization}
\label{fieldtheory}

\subsection{Formalism and exact relations}

Let us now study the equation of motion (\ref{comoving}) using a full
FRG treatment. This will enable us to describe the physics at all
lengthscales and in particular the depinning and creep regime.

A natural framework for computing perturbation theory in
off-equilibrium systems is the dynamical 
formalism\cite{janssen_dynamics_action,martin_siggia_rose}.
After exponentiating the equation of motion (\ref{notcomoving})
using a {\it response field} $\hat {u}$, the average over thermal
noise and disorder can safely be done and yields the simple
(``unshifted'') action
\begin{eqnarray}\label{actionuns}
S_{\rm uns}
(u,\hat{u})&=&\int_{rt} i\hat{u}_{rt}(\eta \partial_t - c \nabla^2)u_{rt}
-\eta T \int_{rt}i\hat{u}_{rt}i\hat{u}_{rt}\nonumber \\
&&-f\int_{rt}i\hat{u}_{rt}\\
&&-\frac{1}{2} \int_{rtt'}
i\hat{u}_{rt}i\hat{u}_{rt'}\Delta(u_{rt}-u_{rt'})\nonumber 
\end{eqnarray}
Disorder and thermal averages $\overline{\langle A[u] \rangle
}=\langle A[u] \rangle_{S_{\rm uns}}$ of any observable  $A[u]$ can be
computed with the weight  
$e^{-S_{\rm uns}}$. 
Furthermore, response functions to an external perturbation $h_{rt}$
added to the right hand side of (\ref{notcomoving}) are simply given by
correlations with the response field: $\langle A[u] i\hat
{u}_{rt}\rangle = \frac{\delta}{\delta h_{rt}}\langle A[u] \rangle $.
It can be checked that causality is satisfied:
$\langle A[\{ u_{t'} \}_{t'>t},\hat {u}] i\hat {u}_{rt}\rangle $ vanishes.
In the time-continuum, the reponse to a perturbation at time $t$ of an
observable depending on $u_{t}$ is ill-defined. We choose Ito
convention for the equation of motion, which ensures that equal time
response functions, and hence any diagram occurring in perturbation theory
containing a loop of response functions, vanish.
The continuum field theory necessarily breaks down at small scales and
it becomes necessary to cut off the integrals over the modes at large
$q$, using a large wave vector $\Lambda $. A full summary of the
notations can be found in Appendix~\ref{app:notations}.

It proves more convenient to work in the {\it comoving frame} 
(i.e. with equation
(\ref{comoving})). The corresponding action is
\begin{eqnarray}\label{eq:action}
S(u,\hat{u})&=&\int_{rt} i\hat{u}_{rt}(\eta \partial_t - c \nabla^2)u_{rt}
-\eta T \int_{rt}i\hat{u}_{rt}i\hat{u}_{rt}\nonumber \\
&&-\tilde{f}\int_{rt}i\hat{u}_{rt}\\
&&-\frac{1}{2} \int_{rtt'}
i\hat{u}_{rt}i\hat{u}_{rt'}\Delta(u_{rt}-u_{rt'}+v(t-t'))\nonumber
\end{eqnarray}
where the field $u$ satisfies $\overline{\langle
\partial_{t}u_{rt}\rangle}=\langle
\partial_{t}u_{rt}\rangle_{S}=0$. This condition fixes
$\tilde{f}\equiv f-\eta v$ in (\ref{eq:action}).
This quantity is the (macroscopic) pinning force,
since it shifts the viscous law $f=\eta v$ by the amount of $\tilde{f}$.

Several exact relations can be derived directly from (\ref{eq:action}).
For any static field $h_r$ (vanishing at infinity)
\begin{eqnarray}
S(u+\frac{1}{c}\nabla^{-2}h,\hat{u})=S(u,\hat{u})-\int_{rt}
i\hat{u}_{rt}h_r \nonumber
\end{eqnarray}
Performing the change of variable $u\rightarrow
u+\frac{1}{c}\nabla^{-2}h$ gives
\begin{eqnarray}
&&\int Du\,\,D\hat{u}\,\,u_{rt}\,\,e^{-S(u,\hat{u})}=\nonumber \\
&&\int
Du\,\,D\hat{u}\,\,(u_{rt}+\frac{1}{c}\nabla^{-2}h_r)\,\,e^{-S(u,\hat{u}) 
+\int_{rt} i\hat{u}_{rt}h_r}\nonumber
\end{eqnarray}
Applying $\frac{\delta }{\delta h_r}|_{h=0}$ yields the exact relation
\begin{equation}
\int_t {\cal R}_{qt}=\frac{1}{cq^{2}}
\end{equation}
where we denote by ${\cal  R}_{rt}$ the exact response function. This
symmetry, known as statistical tilt symmetry,
ensures that the elasticity is {\it not} corrected during the
renormalization.

Another important relation can be derived from
\begin{equation}
\frac{d}{df}\langle
\partial_{t}u_{rt}\rangle_{S_{\rm uns}}=\int_{r't'}\partial_{t}\langle
u_{rt}i\hat {u}_{r't'}\rangle_{S_{\rm uns}}
\end{equation}
This leads to the identity between the macroscopic mobility and the slope
of the $v$-$f$ characteristics at any drive and any temperature:
\begin{eqnarray} \label{exactslope}
\frac{d}{df} v(f) =
\lim_{\omega \rightarrow 0}{-i\omega {\cal R}_{q=0,\omega}}\label{dvdfomrom}
\end{eqnarray}
This exact result can also checked explicitely in the case of 
a particle moving
in a one-dimensional environment \cite{chauve_particle_1d}.

To extract the physical properties from the action (\ref{eq:action}) it
is necessary to build a perturbative approach in the disorder. A
particularly simple case\cite{noteonkpz} occurs when the velocity is 
very large. In that case the disorder operator in the action can be
formally replaced by
\begin{equation}
-\frac{1}{2}\int_{rtt' }i\hat {u}_{rt}i\hat {u}_{rt'}\Delta (v (t-t'))
\end{equation}
since one may neglect the $u_{rt}-u_{rt'}$ compared to $v (t-t')$.
This trick suppresses the non-linearity and the remaining action is quadratic.
Furthermore, at large velocity, $\Delta (v (t-t'))$ can be replaced by
$\delta (v (t-t' ))\int \Delta = \frac{1}{v}\delta (t-t')\int \Delta $ and the
disorder operator transforms into a temperature operator (because it
becomes local in time $t=t'$). The
resulting action is the dynamical action associated to the
Edwards-Wilkinson equation 
\cite{edwards_wilkinson} 
describing the motion of an elastic system in a purely thermal noise
\begin{equation}
\eta \partial_{t}u_{rt} = c\nabla^{2} u_{rt} + \nu_{rt}
\end{equation}
with $\langle \nu _{rt} \nu _{r't'}\rangle =2\eta (T+T_{\rm ew})\delta _{rr'}
\delta _{tt'}$, Langevin noise \cite{noteonkpz} of additional temperature 
$T_{\rm ew}=\frac{\int\Delta}{2\eta v}$ .

Note that at $T=0$ the results at large $v$ coincide with the
pertubative expansion in powers of 
the disorder. The equal-time correlation function in the driven system with
force $f$ crosses over from the static
$\frac{1}{q^{4}}$ Larkin behavior at small scale to a thermal
$\frac{1}{q^{2}}$ behavior at larger scale
\begin{eqnarray} \label{ew}
\overline{u_{-q,t} u_{q,t}}\simeq\left\{
\begin{array}{lll}
\frac{\Delta (0)}{(cq^{2})^{2}}&{\rm for }&q^{2}\gg \frac{\eta v}{cr_{f}}\\
\frac{T_{\rm ew}}{cq^{2}}&{\rm for }&q^{2}\ll \frac{\eta v}{cr_{f}}
\end{array}\right.
\end{eqnarray}
with the same $T_{\rm ew}$, generated at
lengthscales $r\gg \sqrt{\frac{cr_{f}}{\eta v}}$.

\subsection{Renormalization}
\label{renormalization}

We renormalize the theory using Wilson's momentum-shell method. As the cutoff
$\Lambda _{l}=\Lambda e^{-l}$ is reduced, corresponding to a growing
microscopic scale $R_{l}=e^{l}/\Lambda $ in real space, the parameters of the
effective action for slow fields (whose modes $q$ are smaller than
$\Lambda _{l}$) are computed by integration over the fast part of the
fields (whose modes $q$ lie between $\Lambda _{l}$ and $\Lambda $).
This iterative integration gives rise to {\it flow} equations, better
expressed in terms of the {\it reduced} quantities
\begin{eqnarray}
\tilde{\Delta}_{l}(u)&=&\frac{S_{D}\Lambda_{l} ^{D}}
{(c\Lambda_{l} ^{2} e^{\zeta l})^{2}}\Delta_{l}
(u e^{\zeta l})\nonumber\\
\tilde{T}_{l}&=&\frac{S_{D}\Lambda_{l} ^{D}}
{c\Lambda_{l} ^{2} e^{2\zeta l}} T_{l}\label{tilde}\\
\lambda _{l}&=&\frac{\eta_{l} v}{c\Lambda_{l} ^{2} e^{\zeta l}} \nonumber\\
\tilde{f}_{0}&=&f-\eta_{0} v \nonumber
\end{eqnarray}
where $S_{D}$ is the surface of the unit sphere in $D$ dimensions
divided by $(2 \pi )^{D}$. The exponent $\zeta $
is for the moment arbitrary and will be fixed later so that the
reduced parameters flow next to appropriate fixed points. 
In one case (RB) we will need a $l$-dependent $\zeta $, and it is
understood that everywhere the rescaling factors 
$e^{\zeta l}$ (appearing e.g. in (\ref{tilde})) 
should then be replaced by $\exp \int_{0}^{l}dl'\,\zeta _{l'}$.
The reduced
quantities $\tilde{\Delta },\tilde{T}$ are homogeneous to $u^{2}$ and
$\lambda $ to $u$. The parameter $\lambda _{l}$,
which plays a crucial 
role below, can simply be expressed as the following ratio
\begin{eqnarray}\label{lambdavtau}
\frac{v\tau (R)}{\delta u (R)}=\frac{\lambda (R)}{r_{f}}
\end{eqnarray}
of the distance (along $u$) travelled by the
center of mass of the interface during $\tau (R)$ and the roughness
$\delta u (R)=r_{f} (R/R_{c})^{\zeta }$. We have defined $\tau (R)=\eta
(R)R^{2}/c$ as the characteristic relaxation time in the model
renormalized up to scale $R$.

The details of the renormalization procedure can be found in
Appendix~\ref{app:derivation}. The flow equations read:
\begin{eqnarray}
&&\partial \tilde{\Delta}(u)=
(\epsilon-2\zeta ) \tilde{\Delta}(u)
+\zeta u\tilde{\Delta}'(u)
+ \tilde{T} \tilde{\Delta}''(u) \label{flow}\\
&&+ \int_{s>0,s'>0}
\!\!\!\!\!\!\!\!\!\!\!\!\!\!\!\!\!\!\!\!e^{-s-s'}\left[
\tilde{\Delta}''(u)\left( 
\tilde{\Delta}((s'-s)\lambda)-\tilde{\Delta}(u+(s'-s)\lambda)
\right)\right. \nonumber \\
&&-\tilde{\Delta}'(u-s'\lambda)\tilde{\Delta}'(u+s\lambda) \nonumber \\
&&\left.+\tilde{\Delta}'((s'+s)\lambda)\left(
\tilde{\Delta}'(u-s'\lambda)-\tilde{\Delta}'(u+s\lambda)\right)\right]
\nonumber \\
&&\partial \ln \lambda=
2-\zeta -\int_{s>0}e^{-s}s \tilde{\Delta}''(s\lambda)\nonumber \\
&&\partial \ln \tilde{T}=
\epsilon-2-2\zeta +\int_{s>0}e^{-s}s \lambda
\tilde{\Delta}'''(s\lambda)
\nonumber \\
&&\partial \tilde{f}=
e^{ -(2-\zeta )l} c \Lambda_{0}^2 \int_{s>0} e^{-s} \tilde{\Delta}'(s\lambda)
\nonumber
\end{eqnarray}
where $\epsilon =4-D$ and $\partial$ denotes $\frac{\partial}{\partial l}$.

This complicated set of equations require a few comments: (i) as for
the statics \cite{fisher_functional_rg} it is necessary 
to renormalize the whole function $\Delta$, instead of just keeping
few couplings as in standard field theory, (ii) 
the elasticity $c$ is
not renormalized $\partial c=0$
due to the statistical tilt symmetry; (iii) our
equations correctly show that no temperature can be generated at $v=0$
since the fluctuation dissipation theorem holds at equilibrium.

Setting both $T=0$ and $v=0$ in (\ref{flow}) gives back the
simplified set of equations used in
Refs.~\onlinecite{nattermann_stepanow_depinning,narayan_fisher_cdw}
(setting only $v=0$ also yields equations found in
Ref.~\onlinecite{balents_loc}).
But compared to the previous FRG approaches of the depinning
transition, our equations correctly take into account the effect of
the velocity on the flow itself (instead of being treated simply as a
cutoff as in Ref.~\onlinecite{nattermann_stepanow_depinning}). 
Other attempts\cite{stepanow_unpublished} to incorporate velocity and
temperature in the FRG
equations did not obtain the first equation giving
the renormalization of the disorder at $T>0$ and $v>0$. 
To be able to 
tackle the full dynamical problem and study the depinning and the
creep regime, one cannot avoid keeping track of the velocity and of the
temperature in the flow, as will become clear later, since they yield
non-trivial effects which are unreachable by simple scaling arguments.

Our flow equations allow in principle to compute the whole $v$-$f$
characteristics at low temperature. In the following we analyse them in
the three regimes corresponding to the statics ($v=0$), to the
depinning at zero temperature ($T=0$, $f\sim f_c$) and the creep regime
($T>0$, $f \sim 0$).

\subsection{Statics: the cusp}
\label{statics}

At zero velocity, our approach is a dynamical formulation of the
equilibrium problem. It thus allows to recover the known results
about the statics, avoiding the use of replicas.
The standard derivation of the statics using the FRG consists in
writing a replicated Hamilonian for the elastic system
pinned in a random potential
with correlator $\overline{\left( V (r,u)-V (r',u')\right)^{2}}
=-2\delta^{D} (r-r'){\sf R} (u-u')$. After averaging over $V$ the replicated
action reads\cite{fisher_functional_rg}
\begin{eqnarray} \label{replicaction}
S[\overrightarrow{u}]=\frac{1}{2T}\sum_{a}\int_{r}|\nabla
u^{a}_{r}|^{2}-\frac{1}{2T^{2}}\sum_{ab}\int_{r}{\sf R} (u_{r}^{a}-u_{r}^{b})
\end{eqnarray}
where $a,b$ are the $n$ replica indices. 
Performing an FRG analysis of (\ref{replicaction}) yields for the flow
of ${\sf R}$ and $T$ (remarkably independent of $n$):
\begin{eqnarray} \label{flowreplica}
\partial \tilde{{\sf R}} (u)&=&
(\epsilon-4 \zeta) \tilde{{\sf R}} (u) + \zeta u \tilde{{\sf R}}' (u)+ T \tilde{{\sf R}}'' (u)\\
&&+\frac{1}{2}\tilde{{\sf R}}'' (u)^{2}
-\tilde{{\sf R}}'' (0)\tilde{{\sf R}}'' (u)\nonumber \\
\partial \ln \tilde{T} &=& \epsilon -2-2\zeta \nonumber 
\end{eqnarray}
with
$\tilde{{\sf R}}_{l} (u) = e^{-4\zeta l}\frac{S_{D}\Lambda_{l}
^{D}}{(c\Lambda_{l}^{2})^{2}} {\sf R}_{l} (u e^{\zeta l})$ and
$\tilde{T}_{l}=
e^{-2\zeta l}\frac{S_{D}\Lambda_{l}^{D}}{c\Lambda_{l}^{2}}T_{l} $,
which are the same redefinitions as (\ref{tilde}) with 
the correlator $\Delta$ of the force is related to ${\sf R}$ by
(\ref{deltar}). It is easy to see that (\ref{flowreplica}) coincides with our
equations (\ref{flow}) 
when $v=0$ which read
\begin{eqnarray} 
\partial \tilde{\Delta} (u)&=&(\epsilon -2\zeta )\tilde{\Delta}
(u)+\zeta u\tilde{\Delta}
' (u)+ \label{deltav0} \\
&&
+\tilde{T}
\tilde{\Delta} '' (u) +\tilde{\Delta} '' (u) \left(\tilde{\Delta}
(0)-\tilde{\Delta} (u) \right)-\tilde{\Delta}' (u)^{2}\nonumber \\
\partial \ln \tilde{T} &=& \epsilon -2-2\zeta\nonumber 
\end{eqnarray}
Thus the two methods 
give the same results for the 
static and equal-time 
physical quantities. The additional information conveyed by the flow
of the friction $\eta $ in the dynamical formalism is discussed later in
\ref{depinninglaw} and
\ref{depinningdiscussion}.

The temperature in the static system is an irrelevant operator, since it
decreases exponentially fast with $l$. 
One thus commonly restricts to the $T=0$ version of the above
equations. In that case, as is obvious from the closed equation
\begin{equation} \label{closed}
\partial \tilde{\Delta }'' (0)=\epsilon \tilde{\Delta }'' (0)-3
\tilde{\Delta }'' (0)^{2}
\end{equation}
the curvature $\Delta ''(0)<0$ (see Figure~\ref{rbrfsketch})
of the correlator, 
for any initial condition, {\it blows up} at a finite length scale for $D<4$
\begin{equation}
l_{c}=\frac{1}{\epsilon }\ln
\left(1+\frac{\epsilon }{3|\tilde{\Delta }''_{0} (0)|} \right)
\end{equation}
which corresponds to
\begin{equation} \label{rcfrg}
R_{c}=e^{l_{c}}/\Lambda \simeq
\left(\epsilon \frac{c^{2}}{3S_{D}|\Delta_{0} '' (0)|} \right)^{1/\epsilon
}\sim
\left(\epsilon \frac{c^{2}r_{f}^{2}}{S_{D}\Delta_{0}  (0)}
\right)^{1/\epsilon }
\end{equation}
when approximating $|\Delta_{0} '' (0)|$ by $\Delta_{0}(0)/r_{f}^{2}$.
One thus recovers the Larkin length (\ref{larkinlength2}).
The blowup of the curvature of $\Delta$ corresponds to the
generation of a {\it cusp} singularity: $\Delta$ becomes non-analytic
at the origin and acquires for $l>l_{c}$ a non-zero $\Delta' (0^{+})<0$.
However, the flow equation for the running non-analytic correlator
{\it still makes sense}. The non-analyticity just signals
the occurence of metastable states.
A well-defined fixed point function
${\sf R}^{*} (u)$ exists for each of the RB, RF, RP cases
when a suitable $\zeta$ is chosen.

In the RP case, $\zeta=\zeta_{\rm eq}=0$ so as to
conserve the period $a$, 
and the fixed point is given by\cite{bragg_glass_global}
\begin{equation}
\Delta^{*} (ax)=\frac{\epsilon a^{2}}{6}\left( \frac{1}{6}-x (1-x)\right)
\end{equation}
for $x\in [0,1)$.

In the RF case, $\zeta=\zeta_{\rm eq}=\epsilon/3$ so as to conserve
the  RF strength
$\int \Delta$ and the fixed point is given by\cite{fisher_functional_rg}
\begin{eqnarray}
\frac{x^{2}}{2}=y-1-\ln y
\end{eqnarray}
where $y\equiv \Delta^{*} (u)/\Delta^{*} (0)$, $x\equiv u
\sqrt{\epsilon/ (3\Delta^{*} (0))}$ and
$\Delta^{*} (0)\simeq 0.5
\epsilon ^{1/3} (\int \Delta _{0})^{2/3}$ (see \ref{rft}).

In the RB case, it has been shown \cite{fisher_functional_rg} by numerical
integration of the fixed point equation
that $\zeta =\zeta_{\rm eq}\simeq  0.2083 \,\epsilon$ 
yields a physical fixed point, for which no analytical expression is
available.

Despite the irrelevance of the temperature, this operator has important
transient effects during the flow, even if we are left
asymptotically with the $T=0$ cuspy fixed point. It can be shown
(see Appendix~\ref{app:temperature}) that the temperature hinders the flow
from becoming 
singular at a finite scale. The running correlator evolves smoothly towards
its cuspy fixed point and remains analytic, as was also noticed in
Ref.~\onlinecite{balents_loc}. As shown in
Appendix~\ref{app:temperature},
the rounding due to temperature occurs in a boundary layer of width
proportional to $\tilde{T}$ around the origin. 
This is confirmed by the existence of a well-defined expansion in 
$T$ (see Appendix~\ref{nextt}). This effect is missed by simple
perturbation theory that would naively suggest that the rounding occurs
on a width proportional to $\sqrt{\tilde{T}}$. Indeed the correlation
function is proportional to $T$ and smoothes $\Delta $ by $\Delta _{\kappa
}\rightarrow \Delta _{\kappa }e^{-\tilde{T}\kappa ^{2}}$. Although not crucial
for the statics this rounding has drastic consequences for the creep as
analysed in Section~\ref{creep}.

Let us return on the differences between the static and dynamical
formalisms. Within the static approach (\ref{flowreplica}) in the  
$T\rightarrow 0$ limit, despite the occurence of the cusp at $l_{c}$,
the RG equation for ${\sf R}_{l} (u)$ still makes sense after $l_{c}$ 
and flows to a fixed point controlled by $\epsilon =4-D$. However the
physical meaning of the cusp is delicate\cite{balents_rsb_frg}. On the
other hand, the use of the dynamical formalism allows to put $T=0$ from the
beginning but adds to the problem a time dimension and the
corresponding parameter, the friction $\eta$. In this dynamical
version of the problem, the cusp has strong physical consequences
which are more immediate: after $l_{c}$, the cusp generates infinite
corrections to the friction. This feature marks the onset of a
non-zero threshold force at scales larger than the Larkin length and
signals that an infinite time is needed to go from one metastable
state to another. Metastability thus appears very clearly in the
dynamical formulation of the statics problem.

A simple physical picture of the cusp in the statics at $T=0$ was given in 
Ref.~\onlinecite{balents_rsb_frg}.
The renormalized potential $V_{\rm ren} (r,u)$ at scales
$R>R_{c}$ develops ``shocks'' (i.e. discontinuities of the force
$-\partial_{u} V_{\rm ren} (r,u)$ of typical magnitude 
$f_{\rm disc} (R)$ at random positions). Let us now extend this description 
to draw the link with the critical force and to include thermal effects.

The force correlator for small $u-u'$ is dominated by the
configurations with a shock present between $u$ and $u'$:
\begin{eqnarray}
\overline{\left(F_{\rm ren} (r,u)-F_{\rm ren} (r,u')
\right)^{2}}\,\sim \, f_{\rm
disc} (R)^{2} \, \frac{dp}{du}|u-u'|
\end{eqnarray}
where $dp/du$  denotes the probability to find a shock between $u$ and
$u+du$. Identifying the rhs with $R^{-D} |\Delta_{\rm ren} ' (0^{+}) (u-u')|$
one finds, using the rescalings (\ref{tilde}), that the discontinuity in
the force has the following scale dependence
\begin{eqnarray}
f_{\rm disc} (R)\sim f_{c} \left(\frac{R}{R_{c}}
\right)^{-(2-\zeta)}\equiv f_{c}^{\rm eff} (R)
\end{eqnarray}
and can thus be identified with an ``effective critical force''
$f_{c}^{\rm eff} (R)$ at scale $R$, which will play a role in the
following (see Subsection~\ref{creepdiscussion}). 
At $R=R_{c}$, $f_{c}^{\rm eff} (R)$ reduces to the true
critical force $f_{c}$.

The renormalized problem at scale $R$ being the one of an
interface in a potential $V_{\rm ren} (r,u)$ with the above
characteristics, one can now easily understand the result that the
cusp of $\tilde{\Delta}_{l} (u)$ is
rounded on a width $\tilde{T}_{l}/\chi $ at $T>0$. Extending the
previous argument, one expects a rounding of a shock
if the barrier between $u$ and $u'$ is of order $T$. Since near a
shock the potential is linear of slope $f_{\rm disc} (R)$, the barrier
is $f_{\rm disc} (R)|u-u'|$, and the thermal rounding should thus occur in a
boundary layer of width $u$ given by
\begin{eqnarray}
f_{\rm disc} (R)\, u \, R^{D} \, \sim \, T
\end{eqnarray}
Using the rescalings (\ref{tilde}), this is indeed equivalent to 
the expression $\tilde{T}_{l}/\chi$ for the width of the boundary
layer in rescaled variables found in Appendix~\ref{app:temperature}.

\section{Depinning}
\label{depinning}

At $T=0$ and $v\rightarrow 0$, our flow equations give a
self-contained picture of the depinning transition. Thanks to our
formalism, the problem is reduced to the mathematical study of
(\ref{flow}), which although complicated, requires no additional
physical assumptions. To focus on the depinning transition, we must
analyze the solutions of these equations in the regime of small
velocitiy where, using (\ref{tilde}), $\lambda_{l=0}$ is small. We
will examine the various regimes in the RG flow keeping in mind that
$\lambda _{l}$ increases monotonically with $l$.

The equations (\ref{flow}) involve averages over a range $u\sim \lambda
_{l}$ and thus one naturally expects that, at least at the beginning
of the flow, $\Delta _{l} (u)$ remains close to the $v=0$ solution. The two
functions will differ in a boundary layer around $u=0$ of width
denoted by $\rho_{l}$. Although the precise form of the solution for
$|u|<\rho _{l}$ (e.g. whether the cusp persists at $v>0$) is very hard
to obtain analytically, fortunately most of our results will not
depend on such details. As we discuss below, the main issue will be to
decide whether $\rho_{l} \ll \lambda_{l} $ or not, which is a
well-posed mathematical question. 

Let us start by analyzing the flow up to the Larkin scale $l_{c}$ of
the statics, at which the cusp occurs and the corrections to the friction
become singular in the $v=0$ flow.
Here at $v\gtrsim 0$ one enters at $l_{c}$ a regime where 
$\tilde{\Delta}_{l}$ is close to its fixed point 
(see Appendix~\ref{app:before}). Within the boundary layer, the effect
of the velocity is to decrease the singularities of the statics. As shown in 
Appendix~\ref{app:v}, the blow-up of the curvature
$\tilde{\Delta}^{\prime \prime}(0)$
is slowed down by the velocity as
\begin{eqnarray}
\partial \tilde{\Delta}^{\prime \prime}(0)&=&\epsilon
\tilde{\Delta}^{\prime \prime}(0)-3 \tilde{\Delta}^{\prime
\prime}(0)^{2} \nonumber\\ 
&&- 9\lambda^{2}
\tilde{\Delta}^{\prime \prime}(0)\tilde{\Delta}^{\rm iv}(0)+ {\cal O}
(\lambda^{4})\nonumber
\end{eqnarray}
and the same is true for the friction
\begin{eqnarray}
\partial \ln \eta &=&-\tilde{\Delta}^{\prime \prime}(0) -3 \lambda
^{2}\tilde{\Delta}^{\rm iv}(0) +  {\cal O} (\lambda^{4})\nonumber
\end{eqnarray}

If the blurring of the singularity results in a suppression of the
cusp, i.e. if $\tilde{\Delta }_{l}$ remains analytic, one should
wonder whether
the $v\simeq 0$ flow can really {\it remain} close to $\Delta ^{*}$ since
the convergence to the fixed point is crucially dependent
on the existence of the non-analyticity and in particular on the term
$-\tilde{\Delta}' (0^{+})^{2}$ 
in the flow of $\tilde{\Delta}(0)$ in (\ref{deltav0}). A hint that
$\tilde{\Delta}_{l} $ can stabilize for a while at $v>0$ is obtained
by noting that one has (see (\ref{slowdel0}))
\begin{eqnarray}
\partial \tilde{\Delta} (0)&=&(\epsilon -2\zeta )\tilde{\Delta}
(0)
\nonumber\\
&&-\int_{s>0}\!\!\!\!\!\!e^{-s}\left(
\int_{s'>0}\!\!\!\!\!\!e^{-s'}\left(\frac{\tilde{\Delta} (\lambda s)-
\tilde{\Delta} (\lambda s') }{\lambda } \right) \right)^{2}\nonumber
\end{eqnarray}
which has indeed the correct sign to give
the same effect.

Hence it is natural to expect for $l>l_{c}$ that $\tilde{\Delta}_{l}(u)$
has reached everywhere a fixed point form except in the boundary layer. 
The correction to the friction, crucial to determine the $v$--$f$
characteristics, reads
\begin{eqnarray}\label{etaflow}
\partial_{l}\ln \eta _{l}=-\int_{s>0}e^{-s}s\tilde{\Delta} _{l}''
(s\lambda _{l})
\end{eqnarray}
and thus depends on the values of $\tilde{\Delta} _{l}(u) $ for $u\sim
\lambda _{l}$. To estimate this expression, one must know whether the
width $\rho _{l}$ of the boundary layer is smaller than $\lambda _{l}$ or
not.

To summarize these preliminary remarks, the flow in the 
{\it Larkin regime} $l<l_{c}$ is similar to the $v=0$ flow and
$\tilde{\Delta}_{l\gtrsim l_{c}}$ is close to $\Delta^{*}$ except for 
$|u|<\rho _{l}$. We will now analyze in details the flow for $l>l_{c}$
under the assumption that
\begin{equation}\label{rho}
\rho _{l}\ll \lambda _{l}
\end{equation}
As mentionned above, the validity of (\ref{rho}) can in principle be
established by a mathematical or a numerical analysis of our
equations. It turns out that (\ref{rho}) 
leads to the most physically reasonable results. The alternative case
will be discussed below.

\subsection{Derivation of the depinning law}
\label{depinninglaw}

For $l>l_{c}$, called the {\it depinning regime}, and 
relying on (\ref{rho}), the flow of $\eta $ becomes 
\begin{eqnarray}
\partial_{l} \ln \eta _{l}\simeq -\Delta ^{*\prime \prime} (0^{+})
\end{eqnarray}
The friction is renormalized downwards with a non-trivial exponent
$-\Delta ^{*\prime \prime} (0^{+})=-\frac{\epsilon -\zeta }{3}$ with
$\zeta =\frac{\epsilon }{3}$ for the RF case (see Appendix~\ref{rft}) 
and $\zeta =0$ for the RP
case (see Appendix~\ref{rfp}). For the random bond one would naively
take the static $\zeta_{\rm eq}$. 
However our flow equations show that during the
Larkin regime, the form of the disorder correlator evolves to a RF,
and thus $\zeta =\frac{\epsilon}{3}$ also in this case. This non
trivial effect of the transformation for the dynamical properties of a RB
into a RF is discussed in details in Section~\ref{depinningdiscussion}.

Since $\lambda_{l}$ keeps on growing in the depinning regime, the assumption
that $\tilde{\Delta}_{l}(u)$ can be replaced by $\Delta^{*}$
will cease to be valid. This occurs when $\lambda _{l}$ reaches the
range $r_{f}(l)$ 
of $\tilde{\Delta}_{l} (u)$, correlation length of the running disorder.
This defines a scale $l_{V}=\ln \Lambda R_{V}$ 
given by $\lambda_{l_{V}}=r_{f}(l_{V})$.
Above this scale, one enters a regime where the corrections due to disorder
are simply washed out by the velocity,
since the integrals over $s,s'$ in (\ref{flow}) average completely
over the details of $\tilde{\Delta}_{l} (u)$. One thus enters the
{\it Edwards Wilkinson regime}. 
Perturbation theory (\ref{ew}) shows
that the interface is flat for these large scales for $D>2$, the
disorder leading only\cite{noteonkpz} to the effective temperature $T_{\rm
ew}$. 

The family of systems indexed by $0\leq l<\infty $ have all the same
velocity $v$ and the same slope $df/dv (v)$. 
However they have lesser and lesser singular behavior $f(v)$. 
We can thus iterate the FRG flow up to a point where the theory can
solved perturbatively (e.g. above $l_V$). For the
depinning, one can simply use the fact that the renormalized action at
$l=\infty$ is gaussian and its friction $\eta_{\infty}$ is, from
(\ref{exactslope}) equal to the slope $df/dv$ 
of the depinning characteristics. Using the flow of $\lambda _l$ in
(\ref{flow}), the expressions for $\lambda _{l_{V}}$,
$\lambda_{l_{c}}$ with (\ref{tilde}) and (\ref{chirfrf}) lead to 
\begin{equation}\label{flowham}
\frac{\lambda _{l_{V}}}{\lambda _{l_{c}}}\approx \left\{ \begin{array}{l}
\exp \left[(2-\zeta)\beta  (l_{V}-l_{c}) \right]\\
\frac{F_c}{\eta_{l_{c}}v}
\end{array}\right.
\end{equation}
with 
\begin{equation} \label{theexpo}
\beta = 1-\Delta ^{*\prime \prime } (0^{+})/(2-\zeta)
\end{equation}
which will turn to 
be the depinning exponent and we have defined a characteristic force
$F_c=cr_{f}/R_{c}^{2}$. Note that $F_c$ is not exactly the critical force
$f_c$. Solving (\ref{flowham}) gives
\begin{eqnarray}
\frac{R_{V}}{R_{c}}&\approx &\left(\frac{F_c}{\eta _{l_{c}}v}
\right)^{ \frac{1}{(2-\zeta )\beta }} \label{rvrc}\\
\frac{\eta _{l_{V}}}{\eta _{l_{c}}}&\approx&
\left(\frac{\eta _{l_{c}}v}{F_c} \right)^{\frac{1}{\beta }-1}\label{etavetac}
\end{eqnarray}
Since the system at $l_{V}$ is nearly pure, one has
$\eta_{l_{V}}\approx \eta_{\infty }$ and,
integrating over $v$ the derivative $\frac{df}{dv}=\eta _{\infty
}\approx \eta _{l_{V}}$, one gets
\begin{equation} \label{depinexp}
\frac{\eta_{l_{c}}v}{F_c}=\left(\frac{f-f
(v=0^{+})}{F_c} \right)^{ \beta }
\end{equation}
which shows that the depinning is characterized by an
exponent $\beta$ and a pinning force
$f_c = f(v=0^{+})$ (yet to be determined).

The flow of $\tilde{f}_{l}$ allows to fix the value of $f_c$.
Instead of just computing $f_c$ we also show that the integration
of the flow of $\tilde{f}_{l}$ provides a second way to derive the 
depinning law (\ref{depinexp}). Indeed, as discussed below, 
in our formalism the term proportional
to $v$ which was problematic in the previous approaches
\cite{narayan_fisher_cdw} cancels naturally.

In the theory renormalized up to $l_{V}$, the short scale cutoff is
$R_{V}$ and one can use fisrt order perturbation theory. One has
(see (\ref{calvf}) and (\ref{eq:d1}))
\begin{equation} \label{lvperp}
\tilde{f}_{l_{V}}=-\int_{t}\Delta_{l_{V}} '(vt)R^{l_{V}}_{0t}
\end{equation}
Since in the renormalized theory the disorder is close to $\Delta
^{*}$ (with the rescalings (\ref{tilde})), and the friction hidden in the
response function is such that $\lambda_{l_{V}}$ matches
the range of $\tilde{\Delta} _{l_{V}}$, the velocity disappears
from (\ref{lvperp}) which gives
\begin{equation}\label{pertflv}
\tilde{f}_{l_{V}}\simeq e^{ - (2-\zeta ) (l_{V}-l_{c})} A \epsilon
\end{equation}
where $A$ is some constant and the only $v$-dependent quantity is $l_{V}$. 
To connect $\tilde{f}_{l_{V}}$
to the initial parameters, one has to
integrate the flow
$\tilde{f}_{l_{V}}-\tilde{f_{0}}=\int_{0}^{l_{V}}dl\,
\partial_{l}\tilde{f}_{l}$. Expanding $\partial_{l}\tilde{f}_{l}$ in
(\ref{flow}) at small velocity and using
$\tilde{\Delta }' (s\lambda )=\tilde{\Delta }' (0^{+})+ s\lambda
\tilde{\Delta }'' (s\lambda ) + {\cal O} (\lambda ^{2})$ one recognizes
in the second term the correction to $\eta$. Thus for
$l_{c}<l<l_{V}$ one has
\begin{equation}
\partial_{l}\tilde{f}_{l}\simeq e^{ - (2-\zeta )l}c\Lambda
_{0}^{2}\Delta ^{*\prime} (0^{+})-v\partial_{l} \eta_{l}
\end{equation}
where we dropped the sub-dominant terms in velocity. The
integration of the flow gives
\begin{equation}\label{fV-f0}
\tilde{f}_{l_{V}}-\tilde{f}_{0} \simeq -f_{c}\left(1-e^{- (2-\zeta )
(l_{V}-l_{c})} \right)-v (\eta _{l_{V}}-\eta _{0})
\end{equation}
where we defined $f_{c}=c\Lambda _{0}^{2}e^{- (2-\zeta )l_{c}}
|\Delta ^{*\prime} (0^{+}) |/ (2-\zeta )$.
Injecting $\tilde{f}_{0}=f-\eta _{0}v$ and (\ref{pertflv}), we note
that quite remarkably the $\eta _{0}v$ {\it cancel each other}.
We are left with
\begin{equation}\label{f-fc}
f-f_{c} \simeq e^{ - (2-\zeta ) (l_{V}-l_{c})} (A\epsilon -f_{c}) +\eta
_{l_{V}}v
\end{equation}
We already know (\ref{etavetac}) 
that $\eta _{l_{V}}\sim v^{\frac{1}{\beta }-1}$ and (\ref{rvrc})
$e^{l_{V}}\sim v^{ \frac{1}{(2-\zeta)\beta}}$ thus both terms r.h.s
of (\ref{f-fc}) scale like $v^{1/\beta }$.
This leads to the following
result to lowest order in $\epsilon $:
\begin{eqnarray}\label{depinningresult}
f_{c}&\simeq& \left\{\begin{array}{cc}
\frac{\epsilon }{2}\frac{cr_{f}}{R_{c}^{2}}&\mbox{ RF}\\
\frac{\epsilon }{12}\frac{ca}{R_{c}^{2}}&\mbox{ RP}\\
\end{array} \right.\\
v&\sim& (f-f_{c})^{\beta }\\
\beta &=& \left\{\begin{array}{cc}
1-\frac{2\epsilon }{9}&\mbox{ RF}\\
1-\frac{\epsilon}{6}&\mbox{ RP}
\end{array} \right.
\end{eqnarray}
where we used the fact that $\zeta =\epsilon /3$ (RF or RB) or $\zeta =0$
(RP) and the link between $\chi =|\Delta ^{*\prime} (0^{+}) |$ and
$r_{f}$ (RF) or $a$ (RP) stated in Appendix~\ref{app:before}.
In addition, we assumed that $\eta _{l_{c}}$ has a regular behavior when
$v\rightarrow 0$, a non-trivial 
point which we discuss in Subsection~\ref{open}.

\subsection{Discussion}
\label{depinningdiscussion}

The approach of the previous Subsection~\ref{depinninglaw} allows 
us to obtain the characteristics of the $T=0$ depinning. 
We extract the depinning exponent $\beta$, 
the pinning force $f_c$ and the characteristic lengthscales 
from the equation of motion without any additional physical hypothesis or 
scaling relation. Although the depinning problem the exponent 
$\beta$ and the critical force were determined in previous
studies\cite{narayan_fisher_cdw,nattermann_stepanow_depinning}, 
our method is an improvement in several ways.

To get the depinning exponent and critical force, two main derivations
exist in the litterature. One of them extends the static FRG formalism to
the out of equilibrium depinning problem at zero temperature
\cite{narayan_fisher_cdw}, using an ``expansion'' around an
unknown mean-field solution\cite{fisher_depinning_meanfield}. Instead
of directly looking at the renormalized correlator of the
disorder, the method obliges to deal with the time correlation of the
force, $C (v (t-t'))$ in Ref.~\onlinecite{narayan_fisher_cdw}.
This procedure does not allow for a precise enough calculation of the
$v$-$f$ characteristics to demonstrate 
the cancellation of of the $\eta_{0} v$ term (in our
equation (\ref{fV-f0},\ref{f-fc})). In order to obtain a depinning
exponent $\beta $ different from its ``mean-field'' value $\beta =1$,
it is necessary in Ref.~\onlinecite{narayan_fisher_cdw} to
neglect {\it by hand} in the small $v$ limit a term proportionnal to $v$
against a term proportionnal to $v^{1/\beta}$ with $\beta <1$. Our
method, that directly uses averaging over the disorder and properly
takes into account the velocity in the flow of the renormalized
action, allows to show explicitly the needed cancellation.

The other analytical study
\cite{nattermann_stepanow_depinning,leschhorn_depinning2}
of depinning does not consider the renormalization before the Larkin length
and assumes that the singularity is fully developped beyond this
lengthscale. This amounts to take as a starting point the equation of
motion {\it at zero velocity} with a cuspy correlator for the force,
and the Larkin length as the microscopic cutoff. Since the
anomalous exponent of the friction is
$-\Delta^{*\prime \prime } (0)$ which is ill defined for a cuspy
correlator, one is forced in
this method to argue that it should be replaced by $-\Delta ^{*\prime
\prime }(0^{+})$ which is finite. This prescription and scaling
relations linking the roughness, the depinning and the time exponent
$2-\Delta^{*\prime \prime } (0^{+})$, allows to extract the depinning
exponent. 
In our method, the ambiguities that existed in 
Ref.~\onlinecite{nattermann_stepanow_depinning} 
to write the flow of $\eta$ beyond $l_c$ when using the
zero velocity equations, and the trick $0\rightarrow 0^{+}$
becomes a well-defined mathematical property of our finite velocity RG
equations: if (\ref{rho}) is
confirmed, our approach directly shows that the 
$-\Delta^{*\prime \prime }(0^{+})$ prescription is the correct one and
allows to {\it prove} directly the scaling relations, instead of {\it
assuming} them, to obtain the exponent.

Furthermore the occurence of the asymptotic Edwards Wilkinson regime in
Ref.~\onlinecite{nattermann_stepanow_depinning} has to be put by hand
as a cut in the $v=0$ RG flow. The important correlation length
$R_{V}$ 
(denoted $L_{V}$ in Ref.~\onlinecite{nattermann_stepanow_depinning}) 
at which this regime takes place is thus not
well under control and has to be estimated from dimensional analysis.
In our case the depinning regime is naturally cut when our
$\lambda $, which tells how fast the system runs on the disorder,
reaches the range of the flowing correlator. The scale $R_{V}$ at
which it occurs, and above which the non-linearities are washed out,
can clearly be identified with the correlation length of
the moving interface (or more precisely, of the velocity-velocity
correlation). The physical interpretation of (\ref{rvrc}), i.e.,
\begin{eqnarray}
R_{V}\sim R_{c} \left( \frac{f_{c}}{\eta v}\right)^{\frac{1}{z-\zeta }}
\end{eqnarray}
with $z=2-\frac{\epsilon -\zeta }{3}$ is the following: $R_{V}$ 
is the scale at which ``avalanches'' occur in the driven deterministic
sytem. The motion proceeds in a succession of such processes, where
pieces of interface of typical size $R_{V}$ depin over a distance $r_{f}
(R_{V}/R_{c})^{\zeta }$ during a time $r_{f} (R_{V}/R_{c})^{\zeta }/v$.

In addition to providing a clean derivation of the depinning exponents
and of the critical force, our equations contain new physics that was
unreachable by the previous methods.

Although in principle one would expect three universality classes (RF,RB,RP)
for the depinning exponent, it was conjectured by Narayan and Fisher
\cite{narayan_fisher_depinning} that the roughness exponent of the
system at the 
depinning transition for RB or RF
is equal to the roughness exponent of
the static RF case, $\zeta =\epsilon/3$.
This result {\it cannot} be obtained by the approach of Narayan and Fisher
or that of Nattermann {\it et al.} since these authors did not include
the velocity in their RG analysis, and simply treated the small $v$
limit as $v=0$.
On the contrary, our flow equations for the correlator shows directly that a
RB disorder does indeed evolve during the flow towards a RF disorder,
leaving only two different universality classes (RF,RP) for the
dynamics against three for the statics (RB,RF,RP). Such evolution 
is shown on Figure~\ref{rbrfpic}, where an initial RB becomes {\it
dynamically} a RF. In
Appendix~\ref{app:v} we show that the correction to $\int \Delta $,
which measures the RF strength of the disorder, grows as
\begin{eqnarray}
\partial\int \tilde{\Delta} =
2\lambda^{2}\int \tilde{\Delta}^{\prime \prime 2} + {\cal  O}
(\lambda^{4})\nonumber
\end{eqnarray}
where we have used $\zeta=\epsilon /3$. This ensures that a
moving system, even at {\it arbitrary small} velocity, sees an
effective {\it random field} 
at large scale. 
\begin{figure}[htb]
\centerline{\fig{8cm}{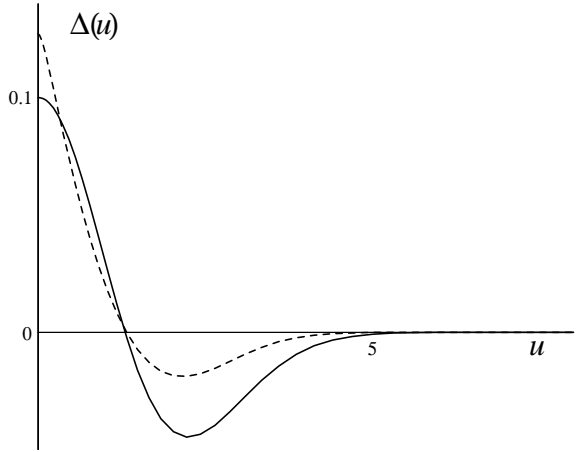}}
\caption{{\narrowtext The evolution of a random bond to a random field
correlator obtained by numerical integration of the flow. The initial condition
$l=0$ of the flow is a RB ($\Delta (u)$ shown as a full line on $u\geq
0$). Following (\ref{flow}), the running correlator transforms into
a RF as shown on the snapshot of $\Delta _{l} (u)$ near $l_{c}$
(dashed line), as can be seen by comparing with the characteristic
shapes of RB/RF shown in Figure~\ref{rbrfsketch}.}}
\label{rbrfpic}
\end{figure}

\subsection{Open questions}
\label{open}

Our FRG equations prompt for several remarks and questions. In the
previous sections, we have examined the consequences of 
the property (\ref{rho}) and established in that case that the values of the
exponents were the ones proposed in 
Refs.~\onlinecite{narayan_fisher_cdw,nattermann_stepanow_depinning,%
narayan_fisher_depinning}.
Although we consider it as unlikely, we have not been able to rule out
the possibility that either $\rho _{l}\sim
\lambda _{l}$ (or even worse, $\rho _{l}> \lambda _{l}$) and thus we should
examine the consequences of a violation of
property (\ref{rho}). If $\rho _{l}\sim \lambda _{l}$, it is not
excluded a priori that there exists another ``fixed point''
behavior (e.g. with a scaling function of $u/\lambda _{l}$). 
However in that case, the exponents should differ from the
standard ones (unless some hidden and rather mysterious sum rule 
would fix the value of the integral in (\ref{etaflow})).
In the absence of an identified fixed point,
it is not clear whether universality would hold.
Again this crucial point (\ref{rho}) can be definitely
answered by an appropriate integration of (\ref{flow}).
Thus the present approach, which clearly takes $v$ into account,
identifies as (\ref{rho}) the condition under which the trick used in 
Refs.~\onlinecite{narayan_fisher_cdw,nattermann_stepanow_depinning}
gives the correct exponents. 

Another intriguing point concerns the continuity between the $v=0$ and
the $v\rightarrow 0$ problems. Indeed, to derive the depinning law
(\ref{depinningresult}) 
we have assumed that $\eta _{l_{c}}$ remains finite as
$v\rightarrow 0$. However, we should recall that in the non-driven
case ($v=0$ and $f=0$), $\eta _{l}$ diverges at $l_{c}$ and thus $\eta
_{l_{c}}=\infty $\cite{etainfini}. If there is any continuity 
in the RG flow as $v\rightarrow 0$ then 
$\eta _{l_{c}}\rightarrow \infty $ in this
limit. In that case the consequence would be (see (\ref{depinexp})) a
modification of the
exponent $\beta \rightarrow \beta / (1-\alpha) $ if $\eta _{l_{c}}\sim
v^{-\alpha }$ (or weaker logarithmic multiplicative corrections). 
We would then find for the
depinning a different result from the conventional one. Since we are 
unable to solve analytically accurately enough the equation for $\eta$
around $l_c$, one should resort to a numerical 
solution of our flow equations (\ref{flow}) to resolve this question. 
Using (\ref{flow}) it is necessary to check 
that $\eta _{l_{c}}$ does not diverge as $v\rightarrow
0$ like a power of $v$ so as to recover the
standard depinning exponent (\ref{theexpo}). 
The question is of particular importance since, if really
a finite-scale behavior, occuring near $R_{c}$, would control the
macroscopic asymptotic behavior, then again one could wonder whether
universality would hold.

Therefore, the description of depinning in terms of a standard
critical phenomenon may be risky. Indeed as clearly appears in our FRG
approach, since the fixed point at $v=0$ is characterized by a {\it
whole function} $\Delta
^{*}$ (i.e. an infinite number of marginal directions in $D=4$)
rather than a single coupling constant (as in usual
critical phenomena) the effect of
an additional relevant perturbation, here the velocity, can be more
complex due the feedback of $v$ itself on the shape of the
function during the flow. This is particularly clear in the RB case
which dynamically tranforms into RF.

\section{Creep}
\label{creep}

We now deal with the non-zero temperature case. 
The system can jump over any energy barrier and overcome the pinning
forces, thus it {\it moves} with $v>0$ for any drive $f>0$ and
never gets pinned. Let us now show how our equations (\ref{flow})
allow to investigate the {\it creep} regime that occurs when the
system moves very slowly with $f\ll f_{c}$, at low temperature.

\subsection{Derivation from FRG}
\label{creeplaw}

\begin{figure}[htb]
\centerline{\fig{8cm}{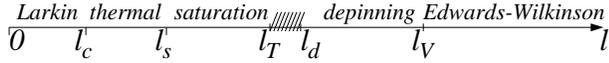}}
\caption{{\narrowtext Characteristic scales and regimes for creep motion.}}
\label{scales}
\end{figure}

As for the depinning, we are interested in infinitesimal
velocities. The bare $\lambda_{0} $ is thus very small. The main 
difference compared to Section~\ref{depinninglaw} is that the 
temperature is now finite as well. The main effect of $T$ is to round 
the cusp in the flow. Since we are interested in extremely small
velocities, we will consider $\lambda_0$ as the smallest quantity 
to start with. A non-zero temperature gives thus rise to a new 
regime in the RG flow, where the rounding of the cusp is due to 
temperature and not to velocity. This leads to the following 
regimes in the FRG flow shown on Figure~\ref{scales}. We
will examine the various regimes in the RG flow keeping in mind that
again, $\lambda _{l}$ increases monotonically with $l$.

Just as in the previous case, we expect a {\it Larkin regime} 
for $0<l<l_{c}$ with small corrections. Above  $l_c$ the disorder
reaches a regime where scaling is imposed by the temperature. Indeed
since $\lambda_{l_c} \ll \tilde{T}_{l_c}/\chi $ one can forget about
the velocity and the  FRG equations are very similar to the $v=0$ and
$T>0$ case. In Appendix~\ref{app:temperature} we show that the
temperature rounds the cusp on a boundary layer $u\sim
\tilde{T}_{l}/\chi $ and we obtain the explicit scaling form
(\ref{scalingform})
\begin{eqnarray}\label{scalingform2}
\tilde{\Delta}_{l} (u)&\simeq& \tilde{\Delta}_{l} (0)-\tilde{T}_{l}f
(u \chi /\tilde{T}_{l}) \nonumber \\
f (x)&=&\sqrt{1+x^{2}}-1 \\
\chi &=& |\Delta^{*\prime}(0^{+})| \nonumber
\end{eqnarray}
which in the statics holds at all scales larger than a scale of order
$l_{c}$. Here, because we focus on $v\rightarrow 0$, the
scanning scale $\lambda _{l_{c}}$ is smaller than the width of the
boundary layer, and the flow of the friction reads in this regime
\begin{eqnarray}\label{chi2surt}
\partial \ln \eta_{l} \simeq -
\tilde{\Delta }''_{l}(0) \simeq \frac{\chi ^{2}}{\tilde{T}_{l} }
\end{eqnarray}
The temperature being irrelevant by power counting,
the initial flow of $\tilde{T}$ is
\begin{equation}\label{ttheta}
\partial \ln \tilde{T}=-\theta 
\end{equation}
since the anomalous correction to $\tilde{T}$
vanish as $\lambda \rightarrow 0$. Here and in the following, 
$\theta=D-2+2\zeta_{\rm eq}$ denotes the energy
fluctuation exponent of the {\it static} problem.  
Together with (\ref{chi2surt}) it shows that the
friction grows extremely fast, like $\exp e^{\theta l}$. 
This is the {\it thermal regime} where motion only occurs via
thermal activation over barriers. The velocity is so small that the
center of mass motion is unimportant and the temperature essentially
flows as in the $v=0$ problem. We have determined the flow in its
initial stages, and we now determine the scale at which this behavior
ceases to hold.

The flow equation (\ref{etacorr}) for $\eta _{l}$ together
with the scaling function (\ref{scalingform2}) for $\tilde{\Delta }$
for $u\sim \tilde{T}_{l}/\chi $ shows that (\ref{chi2surt})
holds only until the new scale $l_{T}=\ln \Lambda R_{T}$ defined as
\begin{equation}
\lambda _{l_{T}}\sim T_{l_{T}}/\chi
\end{equation}
For $l<l_{T}$ the temperature remains the main source of
rounding of the cusp. Above that scale one must take 
the velocity into account.

\begin{figure}[htb]
\centerline{\fig{7cm}{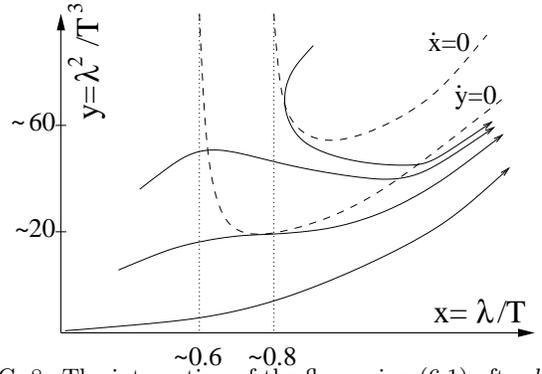}}
\caption{{\narrowtext The integration of the flow using
(\ref{scalingform2}) after $l_{c}$ and reduced variables $x_{l}\equiv
\chi \lambda_{l} /\tilde{T}_{l}$ and $y_{l}\equiv \chi ^{4}\lambda_{l}
^{2}/\tilde{T}_{l}^{3}$. The dotted lines indicate the set of points
where $\partial_{l}x_{l}=0$ or $\partial_{l}y_{l}=0$. Some
trajectories are displayed, with an arrow showing the direction of
growing $l$. The initial conditions for creep are close to the origin,
and closer to the $x$-axis.}}
\label{tsat}
\end{figure}

In fact, this simple picture is not complete since, before reaching
$l_T$ another phenomenon occurs, leading to another lengthscale. 
In the thermal regime the correction to $\tilde{T}$ due to disorder 
competes with the simple exponential decay and (\ref{ttheta}) breaks
down. This physically expresses that motion in a disordered landscape
generates a thermal noise (provided some thermal noise is already
present). Using (\ref{scalingform2}), one has
$\partial \ln  \tilde{T}\approx -\theta +6\chi ^{4}\lambda
^{2}/\tilde{T}^{3}$ at small $\lambda $. Thus the correction to
$\tilde{T}$ reverts at a scale $l_{s}=\ln \Lambda R_{s}$ 
such that $\lambda _{s}\sim
\tilde{T}_{l_s}^{3/2}/\chi ^{2}$. Note that $l_{s}<l_{T}$. 
Above $l_{s}$ the temperature does not decrease any more due to
heating by motion. One can show using (\ref{flow})
that $\tilde{T}$ saturates and does not vary much until the scale
$l_{T}$. We call this intermediate regime $l_{s}<l<l_{T}$ the {\it
saturation regime}. 
We checked it using a numerical integration of the flow in
this regime with the scaling form of the disorder
(\ref{scalingform2}). Analytically, if we suppose that after $l_{s}$,
the correction of $\tilde{T}$ due to disorder dominates $-\theta $,
then one would have in this regime an invariant of the flow
$\partial_{l}\left( \tilde{T}_{l}^{2}-6\chi ^{2}\lambda
_{l}^{2}\right)\approx 0$. If this were true, it is clear that
the flow could {\it never} realize the condition $\lambda _{l}\sim
T_{l}/\chi $, possibility that is excluded on physical basis and 
by the numerics shown in Figure~\ref{tsat}.
Despite the saturation of the temperature, (\ref{chi2surt})
remains true after $l_{s}$. Thus the friction and
$\lambda$ keep on growing and one finally reaches the scale $l_{T}$ at which
the scanning length $\lambda_{l} $ crosses the boundary layer width
$\tilde{T}_{l}/\chi$. 

Above $l_{T}$, a rigorous analytical analysis of (\ref{flow})
becomes difficult. We however expect, since the velocity controls now
the boundary layer, a regime similar to the depinning regime at $T=0$
to occur. Using the same arguments than for the depinning, one 
obtains in that regime
\begin{eqnarray}\label{creepdep}
\partial \ln  \eta &=& -\Delta ^{*\prime \prime} (0^{+}) \\
\partial \ln \tilde{T} &=& 2-D-2\zeta 
\end{eqnarray} 
leading again to a decrease of the temperature, even
slightly accelerated by a negative ${\cal O}
(\epsilon ^{4/3})$ exponent. Let us call $l_{d}$ the 
{\it depinning scale} at which one enters such a 
depinning regime. From the above discussion it is very reasonable 
to expect that one goes directly from the
saturation to the depinning regime, i.e.
$l_{d}-l_{T} \sim {\rm cst}$. However we cannot strictly rule out the 
possibility of an intermediate regime (divergent $l_{d}-l_{T}$  when
$v\rightarrow 0$) during which the correction to the friction goes
smoothly from positive (as in the thermal and saturation regimes) 
to negative values (depinning regime). Again, it would be useful 
to settle this point through a numerical solution of our flow equations.
Note that in the RF and RP cases, the exponent
$\zeta $ and the fixed point $\Delta ^{*}$ 
in (\ref{creepdep}) are the same as in the statics. However in the RB
case, one have used a $l$-dependent $\zeta $ which crosses over between
$\zeta _{\rm eq}$ for $l<l_{T}$ and $\zeta =\epsilon /3$ for
$l>l_{d}$ corresponding to the change from RB to RF fixed points 
$\Delta ^{*}$. 

In the depinning regime, motion now proceeds in a similar way than for
the one studied in Section~\ref{depinning}. Here again at large enough scale,
velocity will wash out the disorder for $l>l_{V}$ with $l_{V}$  determined by
$\lambda_{l} \sim r_{f} (l)$. One then enters the Edwards-Wilkinson regime.

Let us now compute from the flow (\ref{flow}) the lenghtscales defined above 
(see Figure~\ref{scales}).

In the thermal regime $l_{c}<l<l_{s}$ one can compute $\frac{\lambda
_{l_{s}}}{\lambda _{l_{c}}}$ either by integrating its flow or by
equating the boundary values to their expression. This gives
\begin{eqnarray} \label{r2d2}
\frac{\lambda
_{l_{s}}}{\lambda _{l_{c}}}\approx \left\{\begin{array}{l}
\exp \left[(2-\zeta_{\rm eq} ) (l_{s}-l_{c})+\frac{U_{c}}{T}\left(e^{\theta
(l_{s}-l_{c})}-1 \right) \right]\\
\left( \frac{T}{U_{c}}\right)^{ 3/2}\frac{f_{c}}{\eta _{l_{c}}v}e^{
-\frac{3}{2}\theta (l_{s}-l_{c})}
\end{array} \right. \nonumber
\end{eqnarray}
where we defined $f_{c}\equiv \epsilon \frac{cr_{f}}{R_{c}^{2}}$ and
$U_{c}\equiv \epsilon ^{2}R_{c}^{D}\frac{cr_{f}^{2}}{R_{c}^{2}}$. Expressing
the scales as a function of the velocity leads to
\begin{eqnarray}
\left( \frac{R_{s}}{R_{c}} \right)^{ \theta }&\approx
&\frac{T}{U_{c}}\ln \left[\left( \frac{T}{U_{c}}\right)^{3/2}
\frac{f_{c}}{\eta _{l_{c}}v}\right] \label{rc1}\\
\frac{\eta _{l_{s}}}{\eta _{l_{c}}}&\approx &
\frac{f_{c}}{\eta _{l_{c}}v} \left(\frac{U_{c}}{T} \right)^{ 1/\mu
}\ln \left[\left( \frac{T}{U_{c}}\right)^{3/2}
\frac{f_{c}}{\eta _{l_{c}}v}\right]^{ -\frac{3}{2}-\frac{1}{\mu }}\label{etac1}
\end{eqnarray}
with $\mu \equiv \theta / (2-\zeta_{\rm eq}  )$.

In the saturation regime $l_{s}<l<l_{T}$ we proceed in the same manner
and obtain
\begin{eqnarray}
\frac{\lambda_{l_{T}}}{\lambda _{l_{s}}}\approx \left\{\begin{array}{l}
\exp \left[(2-\zeta_{\rm eq}  -\frac{\chi ^{2}}{\tilde{T}_{l_{s}}})
(l_{T}-l_{s})\right]\\
\left( \frac{U_{c}}{T}\right)^{ 1/2} e^{ \frac{\theta}{2} (l_{s}-l_{c})}
\end{array} \right. \nonumber
\end{eqnarray}
Thus
\begin{eqnarray}
\frac{R_{T}}{R_{s}}&\approx & 1\label{r1T}\\
\frac{\eta _{l_{T}}}{\eta _{l_{s}}} &\approx & \ln \left[\left(
\frac{T}{U_{c}}\right)^{3/2}
\frac{f_{c}}{\eta _{l_{c}}v}\right]^{\frac{1}{2}}\label{eta1T}
\end{eqnarray}

Assuming $l_{d}\sim l_{T}$, the depinning regime $l_{d}<l<l_{V}$ follows
directly and
\begin{eqnarray}
\frac{\lambda_{l_{V}}}{\lambda _{l_{d}}}\approx \left\{\begin{array}{l}
\exp \left[(2-\zeta)\beta (l_{V}-l_{d})\right]\\
\frac{1}{\epsilon } \frac{U_{c}}{T} e^{ \theta (l_{s}-l_{c})}
\end{array} \right. \nonumber
\end{eqnarray}
leads to
\begin{eqnarray}
\frac{R_{V}}{R_{d}} &\approx& \left(\frac{1}{\epsilon } \ln \left[\left(
\frac{T}{U_{c}}\right)^{3/2}
\frac{f_{c}}{\eta _{l_{c}}v}\right]\right)^{\frac{1}{(2-\zeta )\beta
}} \label{rdV}\\
\frac{\eta _{l_{V}}}{\eta _{l_{d}}}&\approx& \left(\frac{1}{\epsilon }
\ln \left[\left(
\frac{T}{U_{c}}\right)^{3/2}
\frac{f_{c}}{\eta _{l_{c}}v}\right]\right)^{1-\frac{1}{\beta
}}\label{etadV} 
\end{eqnarray}
with $\beta \equiv \frac{2-\zeta -\Delta ^{*\prime \prime }
(0^{+})}{2-\zeta }$ the depinning exponent (and $\zeta $ the dynamical
roughness exponent).

We are now in a position to compute the 
characteristics $f(v)$. We fix a small velovity $v$ and
solve the flow equations for $\lambda _{l}$, $\tilde{\Delta} _{l}$ and
$\tilde{T}_{l}$ up to $l_{V}$. This allows to relate $\tilde{f}_{l_V}$
to the unknown $\tilde{f}_{0}$. We can now use the fact 
that at the scale $l_V$, the disorder is essentially washed out and a 
perturbative calculation of $\tilde{f}_{l_{V}}\approx
\tilde{f}_{\infty }=0$ is possible. Solving 
backwards we determine
$\tilde{f}_{0}$, wich is simply $f-\eta v$ where $f$ is the real force
applied on the system and $\eta =\eta _{0}$ the bare friction.

The correction to $\tilde{f}$ can not be neglected during the
depinning regime, thus, using $\tilde{f}_{0}=f-\eta _{0}v$,
$\tilde{f}_{\infty }=0$ and expressing $\int_{0}^{\infty
}dl\,\partial_{l}\tilde{f}_{l}$ one has
\begin{eqnarray} \label{theflowf}
f-\eta _{0}v\approx -\int_{0}^{\infty
}dl\,\partial_{l}\tilde{f}_{l}\approx \frac{c\Lambda _{0}^{2}\chi
}{2-\zeta}e^{ - (2-\zeta_{\rm eq})l_{d}}
\end{eqnarray}
In the thermal regime there is essentially no correction to the flow 
of $\tilde{f}$. Thus (\ref{theflowf}) is controled by the depinning 
regime and one should integrate essentially between $l_d$ and $l_V$.
In fact due to the exponentially decreasing behavior of the integrand in
(\ref{theflowf}) the whole integral depends in fact only of the behavior
{\it at} the scale $l_d$.
Assuming that $l_{d}\sim l_{T}$, using (\ref{rc1},\ref{r1T}), one sees
that $e^{-(2-\zeta_{\rm eq})l_{d}}\ll v$ for $v\rightarrow 0$ and thus one
obtains
\begin{eqnarray}\label{creepformula}
\frac{\eta v}{f_{c}}\approx \exp
\left[-\frac{U_{c}}{T}\left(\frac{f}{f_{c}} \right)^{-\mu } \right]\\
\mu =\frac{D-2+2\zeta _{\rm eq}}{2-\zeta _{\rm eq}}
\end{eqnarray}
The prefactor in front of the exponential cannot be obtained reliably
at this order. Note that for the creep, contrarily to the depinning, the 
possible divergence of $\eta_{l_{c}}$ when $v\rightarrow 0$
(and $T\rightarrow 0$) does not affect the argument of the exponential
but only the prefactor.

\subsection{Alternative method and open questions}
\label{alternative}

For the depinning it was possible to recover the depinning law 
using both the integration of the flow of $\tilde{f}$ and of the 
friction $\eta$ and the relation (\ref{dvdfomrom}). Although 
one can also use in principle this method for the creep it gives poor
results in this case. Indeed contrarily to the derivation involving 
$\tilde{f}$ one needs here the flow of $\eta$ in {\it all} regimes 
including the depinning regime $l > l_d$, where $\eta$ is still 
renormalized. Since the renormalization of $\eta$ goes
from large positive growth (first like $\exp e^{\theta l}$, then
exponentially) in the thermal/saturation regime to 
negative  in the depinning regime 
(where the system accelerates with subdiffusive
$z<2$) a precise knowledge of 
the behavior around $l_T$ would be needed. Unfortunately the 
lack of precise analytical methods available above $R_{T}$
prevents from computing precisely such a crossover. A crude 
estimate of the flow can thus only give
a bound of the exact result. If we use (e.g. in the RF or RP cases) the
estimates of each regime, and the perturbative estimate of $\eta
_{l_{V}}$ in the theory at $l_{V}$:
$\eta_{\infty }\simeq \eta_{l_{V}}+\int_{t}
\Delta ''_{l_{V}} (vt) t R^{l_{V}}_{0t} \sim e^{-(2-\zeta)
(l_{V}-l_{c})}\epsilon $ (it will appear that $\eta_{l_{V}} $ diverges
faster than $e^{-(2-\zeta)
(l_{V}-l_{c})}$ when $v\rightarrow 0$). 
The product of (\ref{etac1},\ref{eta1T},\ref{etadV}) is equal to
$\frac{1}{\eta _{l_{c}}}\frac{df}{dv}$. Integrated from
$0$ to $v$, it yields
\begin{eqnarray}\label{nonarrhenius}
\frac{\eta _{l_{c}}v}{f_{c}}&\approx &
\exp \left[-\left(\frac{U_{c}}{T}\left(\frac{f}{\epsilon
^{\frac{1}{\beta }-1}f_{c}} \right)^{-\mu } \right)^{\frac{1}{1+\mu
\left(\frac{1}{\beta }-1 \right)}} \right]
\end{eqnarray}
One would thus find, using the $\eta$ method a non-Arrhenius law for the 
creep regime. Even if one cannot strictly speaking exclude this result,
as discussed above it is most likely an artefact of the 
approximate integration of the flow, and only 
a lower bound of the barrier height. Indeed compared to the 
integration of the flow of $\tilde{f}$,
this procedure is much more sensitive
to the neglect of the crossover $l_{T}<l<l_{d}$.
A more precise integration of the 
flow would very likely show a compensation 
between the latent growth of the friction during the decrease of
$\partial \ln\eta $ (for $l_{T}<l<l_{d}$) and the reduction
of the friction occuring in the depinning regime $l_{d}<l<l_{V}$.
Note that if $\frac{df}{dv}$ were equal to $\eta_{l_{T}}$ then, one
would recover (\ref{creepformula}). It would be useful to check 
explicitely on a numerical integration of the flow that such a 
cancellation does occur and verify that the $\eta$ method confirms 
also the result (\ref{creepformula}).

We also note that the precise determination of the lenghtscales for
$R>R_{T}$ depend on obtaining an accurate solution of the RG flow
equations. In the previous section, we have obtained the
formulas (\ref{r1T}) and (\ref{rdV}) under some
assumptions about the mathematical form of the solutions of the flow
in the region where $\lambda_l$ and
$\tilde{T_l}$ cross. These assumptions, discussed in the previous
Section, should be checked further, e.g via numerical integration.
Although this should not affect the creep exponent derived above,
the precise determination of these length scales is important to
ascertain the exact value of the scale $R_V$ (i.e the avalanche
scale discussed below).

\subsection{Discussion}
\label{creepdiscussion}

Since our flow equations (\ref{flow}) include finite temperature and 
velocity, they allow  for the first time to treat the regime of slow
motion at finite temperature, directly from (\ref{notcomoving}). 
As for the depinning we derive directly from the equation of motion
the force--velocity law and we obtain new physics.

The first important result is of course the creep formula itself
(\ref{creepformula}). Our method allows to prove the main physical
assumptions, reviewed in Subsection~\ref{sccreep}, needed for the
phenomelogical estimate, namely: (i) the equal scaling 
of the barriers and the valleys; (ii) the fact that velocity is dominated 
by activation over the barriers correctly described by an Arrhenius law. 
In our derivation such law comes directly from the integration of the 
flow equations in the thermal regime; (iii) the fact that one can use 
the static exponents in the calculation of the barriers. This appears 
directly in the formula (\ref{creepformula}) but can also be seen from the 
fact that in the thermal regime the velocity can essentially be ignored 
in the flow equations. We also recover the characteristic lengthscale
predicted by the
phenomenological estimate. Indeed one can identify the scale
(\ref{r2d2}, \ref{r1T})
$R_{T}\sim R_{c} (f/f_{c})^{-1/ (2-\zeta _{\rm eq} )}$ 
as the $R_{\rm opt}$ of Subsection~\ref{sccreep}.

Our equations allow to obtain additional physics
in the very slow velocity regime. In particular, we see that the slow 
motion consists in two separate regimes. At small lengthscales 
$R < R_T$ the motion is controlled by thermal activation over
barriers as would occur at $v=0$. This is the regime described by
the phenomenological theory of the creep. Qualitatively, the main novel
result obtained here is that the thermally activated regime is
followed by a depinning regime, as shown by our equations. This leads 
to the following physical picture:
at the length $R_T$, bundles can depin through thermal activation. 
When they depin they start an avalanche like process, 
reminiscent of the $T=0$ depinning, up to a
scale $R_{V}$. The propagation of the
avalanche proceeds on larger scales in a deterministic way.
Thus one is left with a depinning-like motion, and the
size of the avalanches is determined by the natural cut of the RG
($\lambda =r_{f}$), i.e., the scale at which the propagating avalanche
motion is overcome by the regular motion of the center of mass. One
recovers qualitatively and quantitatively some features of the
$T=0$ case at intermediate scale. The typical nucleus jumps over an 
energy barrier $U_{\rm b}
\sim U_{c} (R_{T}/R_{c})^{\theta }$ resulting in $v\sim \exp
-\frac{U_{c}}{T}\left(\frac{f}{f_{c}} \right)^{-\mu }$.
This jump of a region of size $R_{T}$ initiates an avalanche spreading
over a much larger size $R_{V}$ which we find to be (see
(\ref{rc1},\ref{r1T},\ref{rdV}))
\begin{eqnarray}\label{rvrcrtrc}
\frac{R_{V}}{R_{c}}\sim \left(\frac{U_{c}}{T} \right)^{\frac{\nu}{\beta} }
\left( \frac{R_{T}}{R_{c}}\right)^{1+\frac{\theta
\nu}{\beta}}
\end{eqnarray}
with $\nu =\beta / (z-\zeta )=1/ (2-\zeta )$ and 
$z=2- (\epsilon -\zeta )/3$ the critical exponents of the
depinning, and $\theta $ the energy exponent of the statics. 
Note that the correlation length $R_{V}$ diverges at small
drive and temperature as 
$R_{V}\sim T^{-\sigma }f^{-\lambda }$ with $\sigma=\frac{\nu}{\beta}
=\frac{1}{z-\zeta}$ and
$\lambda =\frac{1}{2-\zeta _{\rm eq}}+\frac{\mu}{z-\zeta}$.

To push the analogy further one can consider that the avalanches
at lengthscales $R>R_T$ are similar to the ones occuring in a 
regular $T=0$ depinning phenomenon  due to an excess driving
force $\left(f-f_{c} \right)_{\rm eff}$.
Considering a minimal block size $R_{T}$ instead of $R_{c}$ for
this ``creepy'' depinning, $R_{V}/R_{T}\sim \left( f-f_{c}
\right)_{\rm eff} ^{-\nu} $, one obtains for this effective excess
force:
\begin{eqnarray}\label{eff}
\frac{\eta_{\rm eff} v}{f_{c}^{\rm eff}} \sim
\left( \frac{f-f_{c}}{f_{c}} \right)_{\rm eff} ^{\beta} \sim
\frac{T}{U_{\rm b}}
\end{eqnarray}
linking the creepy motion at $T>0$ and the threshold depinning at
$T=0$. As explained before, there might be an uncertainty in the value
of the avalanche exponent, which could be changed by a quantity of
${\cal O}(\epsilon)$. To confirm (\ref{rvrcrtrc}), 
one would need to further check the precise
behaviour of the solution of the RG equations for $R>R_T$. 

One can understand qualitatively that the problem at scale $R>R_{T}$
looks like depinning according to (\ref{eff}). 
The tilted barrier (see Subsection~\ref{sccreep})
$E (R,f)= U_{c} (R/R_{c})^{\theta }-f R^{D} r_{f}
(R/R_{c})^{\zeta_{\rm eq} }$ to be overcome in order to move 
a region of size $R$ (all barriers
corresponding to smaller scales having been eliminated), 
vanishes at\cite{r0rd} $R_{0} (f)\gtrsim R_{T}$. For the $T=0$
depinning problem, one can
define a scale dependent effective threshold force $f^{\rm eff}_{c}
(R)\sim f_{c} (R/R_{c})^{- (2-\zeta_{\rm eq} ) }$ such that $E
(R,f^{\rm eff}_{c} )=0$ (also defined in Subsection~\ref{statics}), 
which corresponds to the force needed to depin scales
larger than $R$ (the true threshold $f_{c}=f^{\rm eff}_{c}
(R_{c})$ being controlled in that case by the Larkin length).
A possible scaling derivation of
(\ref{eff}) is obtained by noting that at $T>0$, {\it non-activated} 
motion at scale $R$ occurs when the tilted barrier $E (R,f)$ is of the 
order of $T$. This yields a $T$ dependent effective threshold force
such that
\begin{eqnarray}\label{ftcrff}
\frac{f^{\rm eff}_{c} (R)-f^{\rm eff}_{c} (R,T)}{f^{\rm eff}_{c}
(R)}\sim  \frac{T}{U_{c} (R/R_{c})^{\theta }}
\end{eqnarray}
At $R=R_{0} (f)$, one has $f=f^{\rm eff}_{c}(R)$ and 
(\ref{ftcrff}) is identical to (\ref{eff}) to
zeroth order in $\epsilon $ (i.e. $\beta =1$). In fact, to apply the
above static barrier argument, it might be better to work in
the co-moving frame where the velocity of the interface vanishes. 
This amounts to replace $f$ by $f-\eta v$ in the
previous argument, and $E (R_{0} (f),f) =0$, 
$E (R_{0} (f),f-\eta v) = T$ gives back (\ref{eff}).

The crossover between thermally activated processes and depinning-like
motion can also be recovered by noting that 
the condition $\lambda _{l}\sim \tilde{T}_{l}/\chi $ which appears in
the FRG flow can be rewritten 
as (using (\ref{tilde}) and (\ref{lambdavtau})): 
\begin{eqnarray}
f^{\rm eff}_{c} (R)\, v\tau (R) \, R^{D} \, \sim \, T
\end{eqnarray}
where the lhs is a natural energy scale involved in the 
depinning due to driving effect of the center of mass. If it is much
larger than $T$, 
depinning effects dominate, while if it is smaller, the dynamics is
activated.

Finally many open questions still remain. 
Technically it would be interesting to reconciliate the two 
methods based on $\eta$ and $\tilde{f}$ which proved to be equivalent
for the study of depinning. In fact although the 
two methods should formally agree, the comparison at a given 
order in the RG is more subtle. Indeed $\frac{d}{dv}\delta
\tilde{f} = -\delta \eta $, by integration over $l$ between $0$
and $\infty $ and derivation with respect to $v$, gives back $\eta
_{\infty }=\frac{df}{dv}$ provided that $\tilde{f}_{\infty
}=0$. However, one should notice that in $\frac{d}{dv}\delta
\tilde{f} = -\delta \eta $, the derivative is understood
at fixed parameters {\it at the given scale}. The occurence of this 
hidden dependence in the velocity in the running parameters makes 
the equivalence between both approaches delicate. However the
additional term is of higher order in
disorder. Thus, as pointed out above, it is very likely that 
a careful integration of the flow of $\eta$ should resolve this discrepancy,
but this remains to be explicitely checked.

As for the $T=0$ depinning, the existence of the {\it depinning regime} at
$l_{d}$ depends on the precise form of the boundary layer in the
presence of a velocity.
Note that the alternative scenario discussed in Section~\ref{open},
e.g. whether or not the depinnig regime is universal, 
would not affect the creep exponents, but only the subleading corrections. 

\section{Conclusion}
\label{conclusion}

We examine in this paper the dynamics of disordered elastic 
systems such as interfaces or periodic structures, driven by an external
force. We take into account both the effect of
a finite temperature and of a finite velocity to derive the general 
renormalization group equations describing such systems.
We extract the main features of the analytical solution to these equations
both in the case of the $T=0$ depinning (shown on Figure~\ref{vfinsp})
and in the ``creep'' regime (small applied force $f$ and finite temperature).

Our RG equations, when properly analyzed, allow
to recover the depinning law $v\sim (f-f_c)^\beta$ 
and the depinning exponent $\beta$ also obtained 
by other methods. However, contrarily to previous approaches that needed
additional physical assumptions, such as scaling relations among exponents
or by hand regularisation, our approach is self-contained, 
all quantities being derived directly from the equation of
motion. It thus provides a coherent framework to solve the difficulties 
and ambiguities encountered in the previous analytical 
studies\cite{narayan_fisher_cdw,%
nattermann_stepanow_depinning}.
In addition our method allows to establish the universality classes 
for driven systems. It shows {\it explicitely} that a random bond type 
disorder gives rise close to a random field critical behavior at the
depinning. Thus the dynamics is characterized by only two
universality classes (random field (RF) for interfaces and random
periodic (RP) for periodic systems) instead of three. Since this
phenomenon is an intrinsically dynamical one, it was out of the reach
of the previous analytical approaches that 
used $v=0$ flow equations together with additional physical
prescriptions using e.g. the velocity as a cutoff on the $v=0$ RG flow.

Of course one of the great advantages of the present set of 
RG equations is to allow for the
precise study of the small applied force regime at finite $T$, for which 
up to now, only phenomenological scaling arguments could be given. 
Our FRG study confirms the existence of a creep law at small applied
force
\begin{eqnarray}
\frac{\eta v}{f_{c}}\approx \exp
\left[-\frac{U_{c}}{T}\left(\frac{f}{f_{c}} \right)^{-\mu } \right]
\end{eqnarray}
with a creep exponent related to the static ones
$\mu=(D-2+2\zeta_{\rm eq})/(2-\zeta_{\rm eq})$, 
where $\zeta_{\rm eq}$ is the statics 
roughening exponent. It provides a framework
to demonstrate, directly from the equation of motion, 
the main assumptions used in the phenomenological scaling derivation
of the creep namely: (i) the existence of a single scaling for both 
the barriers and the minima of the energy landscape of the disordered
system; 
(ii) the fact that the motion is characterized by an activation (Arrhenius)
law over a typical barrier.

\begin{figure}[htb]
\centerline{\fig{5cm}{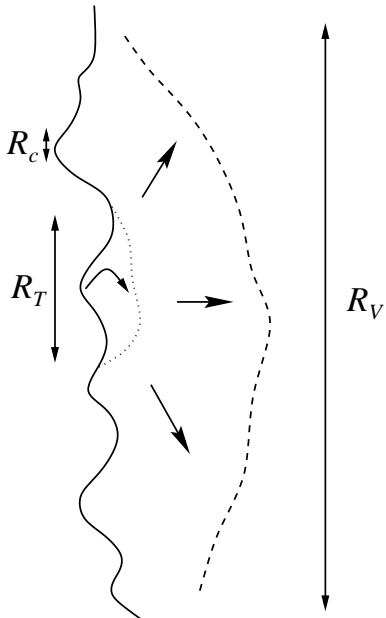}}
\caption{{\narrowtext Schematic picture of the creep process emerging
from the present study: while
thermally activated motion occurs between scales $R_{c}$ (Larkin
length) and $R_{T}$ (thermal nucleus size), depinning-like motion
occurs up to the avalanche size $R_{V}$.}}
\label{dessin}
\end{figure}

In addition, our study unveils a novel ``depinning-like regime'' within the
creep phenomena, not addressed previously, even at the qualitative level 
since the phenomenological creep arguments did not address what happens
{\it after} the thermally activated jump of the optimal nucleus.
Although the velocity
is dominated by the time spent to thermally jump over
the barriers, our equations show that the small $f$ behavior
consists in fact of {\it two different regimes}.
Up to a size $R_T$ motion can only occur through thermal activation 
over barriers.  This is the regime described by the phenomenological
approach to the creep. The optimal nucleus of the scaling estimate 
is given directly by the RG derivation as 
$R_T\sim(1/f)^{1/(2-\zeta_{\rm eq})}$.
Remarkably, another interesting regime exists above this lenghtscale
(see Figure~\ref{dessin}). 
It emerges directly from our RG equations 
and can be given the following simple physical interpretation.
In some regions of the system, bundles of size $R_T$ depin due to thermal
activation. These small events then trigger much larger ones, and the 
motion above $R_T$ proceeds in a {\it deterministic way}, much as the 
$T=0$ depinning. In particular once the initial bundle depins 
it triggers an avalanche up to a size $R_V$ which is given by
$R_V/R_T \sim (U_{c}/T)^{\nu /\beta } (R_T/R_c)^{\theta \nu/\beta}$
where $\theta$, $\beta$ and $\nu $ are the energy, depinning and
correlation length exponents respectively.

The present study also raises several interesting questions which
deserve further investigation, some of them rely on being able
to obtain a more accurate solution of our flow equations. We have shown
explicitly how to recover from our equations the conventional
depinning law (and the scaling creep exponents).
It rested on a mathematical property, likely to hold,
but not yet rigorously established, of the solution for the
flow of the correlator of the disorder. Such behavior should be checked in 
details. The equations being quite complicated,
a numerical solution, albeit delicate, seems to be appropriate.
If the constraint (\ref{rho}) 
on the flow defined in Section~\ref{depinningdiscussion}
were found to be violated, then the conventional picture of the depinning
would very likely fail, as we have analyzed in detail. 
A similar question arises concerning the flow 
of the friction $\eta$ as discussed in Section~\ref{creepdiscussion}.
If the solution of the flow is found to depend on the precise
behavior at the Larkin length $R_c$, it is likely that even universality
could be questioned. These issues are a priori less important for the
first, thermally activated, part of the creep regime, but because
of the existence of a second, depinning-like regime, they would also have
consequences for creep. Again, these question depend on the
precise form of the flow and can be answered unambiguously by a
detailed enough analysis of our equations.
It would also be of great interest to develop a more detailed
physical picture of the crossover between thermally activated and
depinning like motion since we found that both occur within the creep
phenomenon.
   
Several applications and extensions of our work can be envisioned. 
First, extensions to many-dimensional displacement field 
(of dimension $N>1$), given in Appendix~\ref{app:n>1}, would be 
interesting to study within the methods used here. One could check 
whether the approximation used in 
Ref.~\onlinecite{ertas_kardar_anisotropic} yields the correct result
for the $N>1$ depinning. Second, the effect of additional KPZ 
non-linearities could be investigated. In particular one could check
the usual argument which yields that KPZ terms are unimportant for 
the depinning\cite{nattermann_book_young} since their coupling
constant is proportionnal to the (small) velocity. Also, 
extensions to other types of disorder, such as correlated
disorder\cite{chauve_mbog} are possible.
Finally, it should allow to describe in a systematic
way the the thermal 
rounding of the depinning, i.e. the study of 
the $v$--$f$ characteristics for $f$ close to the threshold and small
$T$. If one assumes that one can simply
carry naive perturbation theory in $T$ around the $T=0$ solution of
the RG flow near $f_c$ (i.e. only keeping the contribution beyond $l_V$), 
one is led in (\ref{f-fc}) to an additional term
proportional to $T/v^2$, which readily yields the value for
the thermal rounding exponent $\rho=1+2 \beta$ 
proposed in Ref.~\onlinecite{stepanow_unpublished} 
(i.e. a scaling form near $f=f_c$ and small $T$ for the velocity 
$v\sim T^{\beta/\rho} \Phi\left(\frac{f-f_c}{T^{1/\rho}}
\right)$). Although this exponent seems to be consistent with starting
values $\lambda \ll T$, its 
validity could be further checked by solving our RG flow
equations at small $T$.

\acknowledgements
One of us (TG) would like to thank the Newton Institute (Cambridge)
for support and hospitality.

\appendix

\section{Notations}
\label{app:notations}

Here are some notations and conventions and diagrammatics
we use in the text.

The surface of the
unit sphere in $D$ dimensions divided by $(2\pi)^{D}$ is denoted by
$S_{D}=2 (4\pi )^{-D/2} /\Gamma (D/2)$. 
The thermal average of any observable $A$ is $\langle A \rangle $, the
disorder average is $\overline{A}$, and the average with the dynamical
action $S[u,\hat {u}]$ is denoted by $\langle A \rangle
_{S}=\overline{\langle A\rangle }$.
The Fourier transform of a function $h_{rt}$ of $(r,t)$ is
$h_{q\omega }=\int_{rt}e^{-iq.r+i\omega t}h_{rt}$ where
$\int_{rt}\equiv \int dr\,dt$, and the inversion reads $h_{rt}
=\int_{q\omega}e^{iq.r-i\omega t}h_{q,\omega}$, where
$\int_{q}\equiv \int \frac{d^{D}q}{(2\pi)^{D}}$,
$\int_{\omega}\equiv \int \frac{d\omega}{2\pi}$.
The Fourier transform of the correlator
$\Delta (u)$ is $\Delta_{\kappa}=\int
du\,e^{-i\kappa.u}\Delta (u)$ in general or $\Delta_\kappa=\int_{0}^{a}du\,
e^{-i\kappa u} \Delta(u)$ in the periodic case. One has thus $\Delta
(u)=\int_{\kappa} e^{i\kappa.u}\Delta_{\kappa}$, where
$\int_{\kappa}\equiv \int \frac{d\kappa}{2\pi}$ or
$\frac{1}{a}\sum_{\kappa }$ in the periodic case.
Note that
$\Delta_\kappa$ is a real and even function of $\kappa $.
\begin{figure}[htb]
\centerline{\fig{7cm}{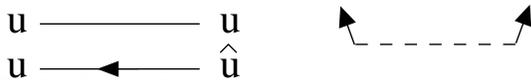}}
\caption{{\narrowtext
Conventions for the diagrammatics. The correlation function $uu$,
which vanishes at $T=0$ is a full line with no arrow.
The reponse $ui\hat {u}$ is a full oriented line.
The vertex $-\frac{1}{2}i\hat
{u}_{rt}i\hat {u}_{rt'}\Delta (u_{rt}-u_{rt'}+v (t-t'))$ is
naturally splitted in two half vertices corresponding to the points
$(r,t)$ and $(r,t')$, and the dashed line means that both points have
the same position. At $T=0$ the correlation vanish.}}
\label{diagfirst}
\end{figure}
The graphs are made of the following units (see Figure~\ref{diagfirst}):
a full line between points $(r,t)$ and $(r',t')$ is a
correlation $\langle u_{rt}u_{r't'}\rangle_{S} $, an {\it oriented line}
with an arrow from point $(r',t')$ to point $(r,t)$ is a
response $\langle u_{rt}i\hat {u}_{r't'}\rangle_{S} $ (the arrow means
that $t>t'$, for the function does not vanish by causality). The
vertex is represented as a dashed line linking points $(r,t)$ and
$(r,t')$. The dashed line means that both points have the same
position $r$. From each point emerges a $\hat {u}$ field.
No arrow is needed for the full line or for the dashed line, since
they are symmetric with respect to the exchange of their endpoints.
The correlation being proportional to $T$ vanishes at $T=0$. The graphs
renormalizing the disorder (see Figure~\ref{diagsecord})
are made of vertices and responses, and
they possess two external $i\hat {u}$ lines. It can
be easily seen that arrows are no more necessary since the two
external $\hat {u}$ lines provide an orientation to all the responses
of the graph. Indeed, due to causality, each of the external $\hat
{u}$ is root of a tree, whose branches are response functions, which
are oriented in the direction of the root.

\section{Perturbation theory}
\label{app:pt}

We derive here the direct perturbation theory at $T>0$ without the use of the
MSR formalism.
To organize the perturbation series, let us multiply the
non-linear part of the equation of motion $F (r,vt+u_{rt})$ by a
fictious small parameter $\alpha$, which will be fixed to one
at the end of the calculation. Directly on
\begin{eqnarray}\label{systemalpha}\left\{
\begin{array}{rcl}
\overline{\langle u_{rt}\rangle }&=&0\\
(\eta \partial_{t}-c\nabla ^{2})u_{rt}&=&
\alpha F (r,vt+u_{rt})+\tilde{f}+\zeta _{rt}+h_{rt}
\end{array} \right.
\end{eqnarray}
we can formally expand $u=\sum_{n\geq 0}\alpha
^{n}u^{(n)}$, $f-\eta v\equiv 
\tilde{f}=\sum_{n\geq 0}\alpha ^{n}\tilde{f}^{(n)}$, 
solve recursively the system (\ref{systemalpha}),
even at non-zero temperature, and compute the
$\alpha $-expansion of every observable. Note
that we added a source $h_{rt}$ (with no constant uniform part)
so as to compute the response function. As the force is
gaussian, the expansion of disorder averaged quantities is in powers
of $\alpha ^{2}$, and is in fact an expansion in powers of $\Delta $.
We denote by ${\cal C}_{r-r',t-t'}=\overline{\langle
u_{rt}u_{r't'}\rangle}$ the exact 
correlation and by ${\cal R}_{r-r',t-t'}=\overline{\langle
\frac{\delta u_{r,t}}{\delta h_{r',t'}}\rangle}$ the exact 
response functions. 

The first iterative steps are $\tilde{f}^{(0)}=\tilde{f}^{(1)}=0$ and
\begin{eqnarray}
(\eta \partial_{t}-c\nabla ^{2})u_{rt}^{(0)}&=&\zeta_{rt}+h_{rt}\nonumber \\
(\eta \partial_{t}-c\nabla ^{2})u_{rt}^{(1)}&=&F
(r,vt+u_{rt}^{(0)})\nonumber \\
(\eta \partial_{t}-c\nabla ^{2})u_{rt}^{(2)}&=&\partial_{u}
F(r,vt+u_{rt}^{(0)})u_{rt}^{(1)}+\tilde{f}^{(2)}\nonumber
\end{eqnarray}
These are sufficient to compute to first order in $\Delta$ the force,
the correlation and response.

In the absence of disorder the system moves with a linear
characteristics $f=\eta v$ and one has the following correlation
and response
\begin{eqnarray}
C_{q\omega}=\frac{2\eta T}{(cq^{2})^{2}+ (\eta \omega)^{2}}
&\qquad&
C_{qt}=T\frac{e^{-cq^{2}|t|/\eta}}{cq^{2}}\\
R_{q\omega}=\frac{1}{cq^{2}-i\eta \omega}
&\qquad& 
R_{qt}=\frac{\theta (t)}{\eta}e^{-cq^{2}t/\eta}
\end{eqnarray}
related by the fluctuation-dissipation theorem (FDT)
$TR_{rt}=-\theta(t) \partial_t C_{rt}$. Note that ${\cal  R}$ and
${\cal  C}$ do {\it not} verify FDT at $v>0$.

\begin{figure}[htb]
\centerline{\fig{8cm}{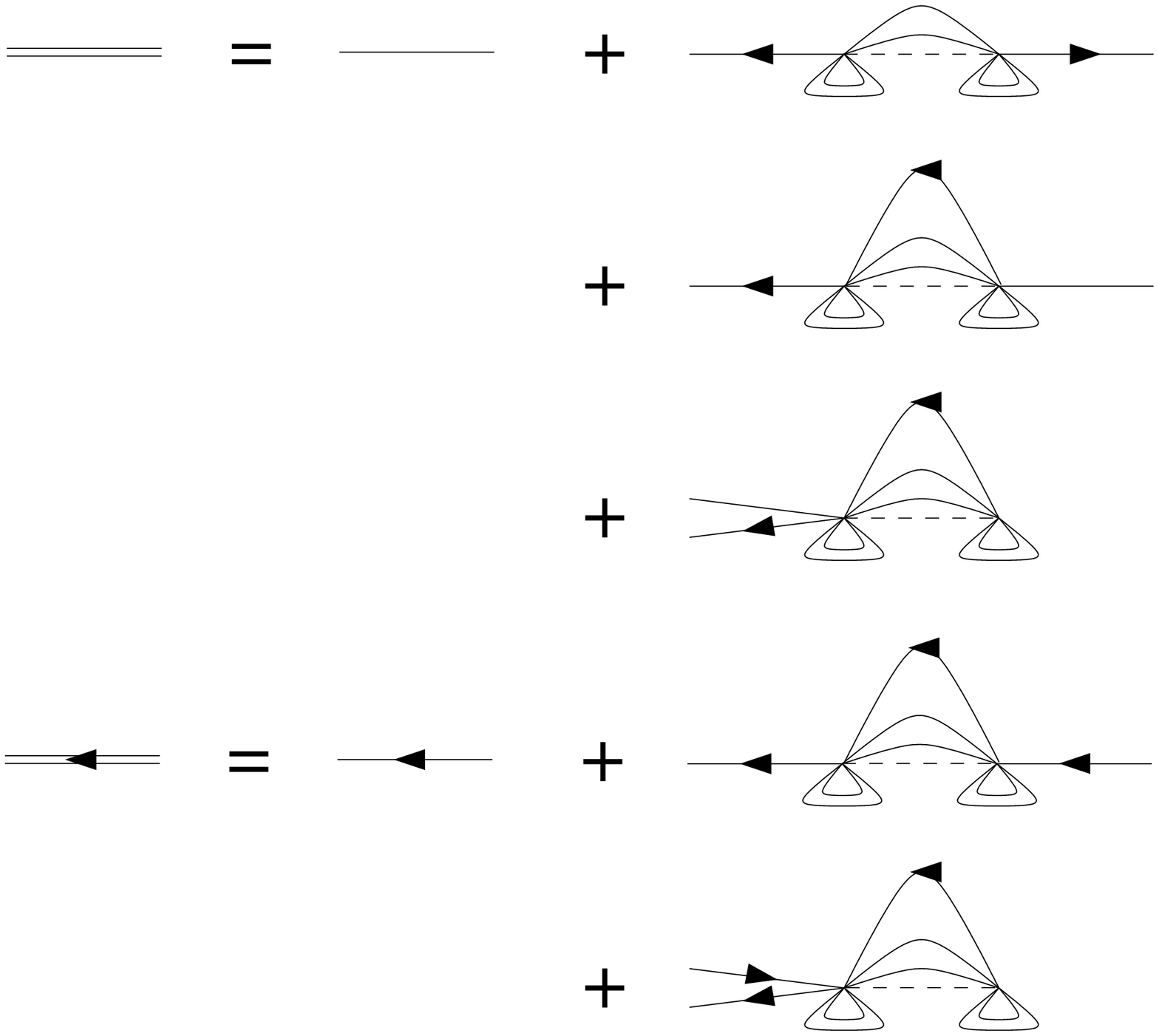}}
\caption{{\narrowtext Computation of the correlation (top) and
response (bottom) functions to first order in perturbation theory. At
$T>0$, the
tadpoles and self-contractions of the vertices contain an arbitrary
number of correlations.}}
\label{corrrespfirst}
\end{figure}

To first order in $\Delta $ one obtains at $T=0$
\begin{eqnarray}
f-\eta v&=&-\int_{\kappa q}\frac{i\kappa
\Delta_{\kappa}}{cq^{2}-i\kappa \eta v}\label{v(f)T=0}\\
{\cal C}_{q,\omega}&=&\frac{\frac{1}{v}\Delta _{\kappa =
-\omega/v }}{(cq^{2})^{2}+ (\eta \omega
)^{2}}\\
{\cal R}_{q,\omega}&=&\frac{1}{(cq^{2})^{2}+ (\eta \omega
)^{2}}\int_{t} \Delta'' (vt)R_{0t}
(1-e^{i\omega t})
\end{eqnarray}
These results can be extended to any temperature $T$:
\begin{eqnarray}
f-\eta v&=&-{\cal D}_{1} (\omega=0)\label{calvf}\\
{\cal C}_{q,\omega}&=&
C_{q,\omega}+\label{calc}\\
&&R_{q,\omega}{\cal D}_{0}(\omega)R_{-q,-\omega}+\nonumber\\
&&R_{q,\omega}\left( {\cal D}_{2} (\omega=0)- {\cal D}_{2}
(\omega)\right)C_{q,\omega}+h.c.\nonumber\\
{\cal R}_{q,\omega}&=&R_{q,\omega}+\label{calr}\\
&&R_{q,\omega}
\left( {\cal D}_{2}(\omega=0)- {\cal D}_{2}(\omega)\right)
R_{q,\omega} \nonumber
\end{eqnarray}
where we have introduced the effective vertices, non-local in time,
smoothed by the temperature (see Figure~\ref{corrrespfirst})
\begin{eqnarray}
{\cal D}_{0}(t)&=&\int_{\kappa}\Delta_{\kappa}e^{i\kappa vt+
(i\kappa)^{2} (C_{00}-C_{0t})}\label{eq:d0}\\
{\cal D}_{1}(t)&=&\int_{\kappa}i\kappa
\Delta_{\kappa}e^{i\kappa vt+
(i\kappa)^{2} (C_{00}-C_{0t})} R_{0t}\label{eq:d1}\\
{\cal D}_{2}(t)&=&\int_{\kappa}(i\kappa)^{2}\Delta_{\kappa}e^{i\kappa
vt+ (i\kappa)^{2} (C_{00}-C_{0t})}R_{0t}\label{eq:d2}
\end{eqnarray}

We now want to compute the
corrections to the parameters $c$, $\eta$, $\tilde{f}$, $T$, $\Delta
(u)$ so that $v$, ${\cal C}_{q\omega}$, ${\cal R}_{q\omega}$ remain
unchanged while the physical (ultra-violet) cutoff
$\Lambda$ on the $q$ integrations is reduced. To first order in
$\Delta$ and $T$, one obtains, 
\begin{eqnarray}
\partial c&=&0\\
\partial \eta
&=&-\int_{t}tR^{>}_{0t}\Delta''(vt)\\
\partial \tilde{f} &=&\int_{t}R^{>}_{0t}\Delta' (vt)\\
\partial T &=&\frac{1}{\eta}\int_{t>0}tC^{>}_{0t}\partial_{t}\Delta''
(vt)\\
\partial \Delta (u) &=& C^{>}_{00} \Delta '' (u)
\end{eqnarray}
with $\partial \equiv -\Lambda \frac{d}{d\Lambda}$ and
$R^{>}_{rt}$, $C^{>}_{rt}$ are the on-shell gaussian
response and correlation functions, i.e., with modes $q$ lying only 
between $\Lambda -d\Lambda $ and $\Lambda $.

A completely different way for obtaining the perturbation expansion
is presented in
Refs.~\onlinecite{chen_marchetti,blatter_vortex_review}, as a first
attempt to include thermal fluctuations in the large-velocity expansion of
Ref.~\onlinecite{larkin_largev}. It consists in
splitting the displacement field into a $T=0$ part and a thermal
part. This procedure is probably only true to first order in
$T$ and not controlled at higher $T$. Instead, the method presented
here is really an expansion in disorder at any $T$.

Although the calculation can in principle be pushed to second order,
the method is too cumbersome to do it in practice (see however 
Ref.~\onlinecite{bucheli_frg_secondorder} at $T=0$). 
It is easier to 
use the formalism of dynamical field theory as shown in
Appendix~\ref{app:derivation}.

\section{Derivation of the flow at finite velocity and finite temperature}
\label{app:derivation}

Here we give the details of the renormalization procedure used for the
moving system. We use the MSR formalism with action $S[u,\hat {u}]$
given by (\ref{eq:action}).
Having shifted the field $u_{rt}$ so that its average
vanishes $\overline{\langle u_{rt}\rangle} =0$, we can do
perturbation theory with the gaussian part
\begin{eqnarray}
S_{0}[u,\hat {u}]=
\int_{rt}\left[ i\hat {u}_{rt}\left(\eta \partial_{t}-c\nabla ^{2}
\right)u_{rt}-\eta T i\hat {u}_{rt}i\hat {u}_{rt}\right]
\end{eqnarray}
of the action.
The gaussian correlation $C_{rt}$ and response $R_{rt}$ functions were
defined in Appendix~\ref{app:pt}.

The interaction part of the action contains the disorder correlator
and also the pinning force $\tilde{f}={\cal O}(\Delta )$:
\begin{eqnarray}
S_{\rm i}[u,\hat {u}]&=&-\tilde{f}\int_{rt} i\hat {u}_{rt}\\
&&-\frac{1}{2}\int_{rtt'}i\hat
{u}_{rt}i\hat {u}_{rt}\Delta (u_{rt}-u_{rt'}+v (t-t')) \nonumber
\end{eqnarray}
The effective action for slow fields $u,\hat {u}$ is given by the
following cumulant expansion where the averages are computed within
the gaussian part $S_{0}$ over the fast fields $u^{>},\hat {u}^{>}$
\begin{eqnarray}
S_{<}[u,\hat {u}]&=&S_{0}[u,\hat {u}]+\langle S_{\rm i}[u+u^{>},\hat
{u}+\hat {u}^{>}] \rangle \nonumber \\
&&-\frac{1}{2}\langle S_{\rm i}[u+u^{>},\hat
{u}+\hat {u}^{>}]^{2} \rangle _{\rm c} + {\cal O} (S_{\rm i}^{3})
\end{eqnarray}
We now turn to the computation of the first and second order terms.

\subsection{First order}

To first order, the corrections arise from the graph shown in
Figure~\ref{fig:diagfo}:
\begin{figure}[htb]
\centerline{\fig{6cm}{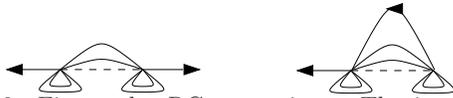}}
\caption{{\narrowtext First order RG corrections. The internal lines
carry fast fields.}}
\label{fig:diagfo}
\end{figure}
They read
\begin{eqnarray}
\left<S_{\rm i}[u+u^{>},\hat{u}+
\hat{u}^{>}]\right>=-\tilde{f}\int_{rt}i\hat{u}_{rt}\nonumber \\
-\int_{rtt'\kappa} (i\kappa )\Delta_{\kappa } [u] (r,t,t')
R^{>}_{0t-t'} i\hat{u}_{rt} \label{firstord}\\
-\frac{1}{2} \int_{rtt'\kappa}\Delta_{\kappa } [u]
(r,t,t') i\hat{u}_{rt}i\hat{u}_{rt'}\nonumber
\end{eqnarray}
with the shorthand notation:
\begin{eqnarray}
\Delta_{\kappa } [u] (r,t,t')\equiv \Delta_\kappa
e^{i\kappa(u_{rt}-u_{rt'} + v(t-t'))} e^{(i\kappa)^2
(C^{>}_{00}-C^{>}_{0t-t'})}\nonumber
\end{eqnarray}
The term (\ref{firstord})
appears to be the sum of a $i\hat{u}{\cal F}[u]$ term and a
$i\hat{u}i\hat{u}{\cal G}[u]$ term. Let us begin to deal with the
first type. A short time expansion of
$e^{i\kappa(u_{rt}-u_{rt'})}$ yields the following operators
\begin{eqnarray}
-(\int_{rt} i\hat{u}_{rt})\int_{\kappa t}i\kappa\Delta_\kappa
e^{i\kappa vt} e^{(i\kappa)^2
(C^{>}_{00}-C^{>}_{0t})} R^{>}_{0t}\label{fcorr}
\end{eqnarray}
which is a correction to $\tilde{f}$ and
\begin{eqnarray}
-(\int_{rt}
i\hat{u}_{rt}\partial_t u_{rt})\int_{\kappa t}(i\kappa)^2
\Delta_\kappa e^{(i\kappa)^2
(C^{>}_{00}-C^{>}_{0t})} tR^{>}_{0t}\label{etacorr}
\end{eqnarray}
which is a correction to $\eta$.
The elasticity operator $i\hat{u}\nabla^2 u$ is {\it not} corrected
and {\it no} higher gradients like
$i\hat{u}\nabla^n u$ are generated in the equation of motion.
Note also that to this order, no KPZ term $i\hat{u} (\nabla
u)^{2}$ is generated\cite{noteonkpz}.

The $i\hat{u}i\hat{u}{\cal G}[u]$ term can be rewritten as the sum of
\begin{eqnarray}
-\frac{1}{2} \int_{rtt'} \!\!\!\!\!\!\! i\hat{u}_{rt}i\hat{u}_{rt'}\int_\kappa
\Delta_\kappa e^{(i\kappa)^2 C^{>}_{00}}
e^{i\kappa(u_{rt}-u_{rt'}+v(t-t'))}\nonumber
\end{eqnarray}
which has the form of a disorder correlator
and yields a correction to $\Delta (u)$, and an operator
quasi local in time
\begin{eqnarray}\label{quasiloc}
&&\int_{rtt'}  i\hat{u}_{rt}i\hat{u}_{rt'} \times \\
&& \times \int_\kappa
\Delta_\kappa e^{(i\kappa)^2 C^{>}_{00}
+i\kappa(u_{rt}-u_{rt'}+v(t-t'))} \left(\frac{1-e^{\kappa^2
C^{>}_{0t-t'}}}{2} \right)\nonumber
\end{eqnarray}
which yields a correction to the
$\int_{rt}i\hat{u}_{rt}i\hat{u}_{rt}$ term. The
projection of (\ref{quasiloc}) on this thermal noise operator is
\begin{eqnarray}
\left(\int_{rt} i\hat{u}_{rt}i\hat{u}_{rt} \right)\int_{\kappa t}
\Delta_\kappa e^{(i\kappa)^2 C^{>}_{00}} e^{i\kappa
vt}\left(\frac{1-e^{\kappa^2 C^{>}_{0t}}}{2} \right)\nonumber
\end{eqnarray}

To obtain the correction to the temperature $T$, one uses
$\frac{\delta T}{T}=\frac{\delta \eta T}{\eta T}-\frac{\delta \eta
}{\eta }$. An integration by parts of (\ref{etacorr}), thanks to FDT
for the ``pure'' $R$ and $C$,
yields $\frac{\delta T}{T}$.

To summarize,
\begin{eqnarray}
\delta c&=&0\nonumber \\
\delta \tilde{f}&=&\int_{\kappa t}i\kappa
\Delta_\kappa e^{(i\kappa)^2 (C^{>}_{00}-C^{>}_{0t})}
e^{i\kappa vt} R^{>}_{0t} \nonumber \\
\delta \eta&=&-\int_{\kappa t>0}(i\kappa)^2
\Delta_\kappa e^{(i\kappa)^2 (C^{>}_{00}-C^{>}_{0t})}
e^{i\kappa vt} t R^{>}_{0t} \nonumber \\
\delta \Delta(u)&=&\int_\kappa \Delta_\kappa e^{(i\kappa)^2
C^{>}_{00}} e^{i\kappa u} \nonumber  \\
\eta \delta T&=&\int_{\kappa t>0} \!\!\!\!\!\!\!
i\kappa vt\Delta_\kappa e^{(i\kappa)^2
C^{>}_{00}}e^{i\kappa vt}\left(1-e^{ \kappa ^{2}C^{>}_{0t}} \right)\nonumber
\end{eqnarray}
The correction $\delta \tilde{f}$ has the same form as the perturbative
expression for $\tilde{f}$, with opposite
sign and shell-restricted functions $C,R$. Note that
$\delta \eta=-\frac{d}{dv}\delta \tilde{f}$.

In the infinitesimal shell limit, the shell-restricted functions
$C^{>},R^{>}$ which
are evaluated at $r=0$ are of order $dl$. The differential flow is
thus given by (\ref{flow}).

\subsection{Second order}

\begin{figure}[htb]
\centerline{ \fig{6cm}{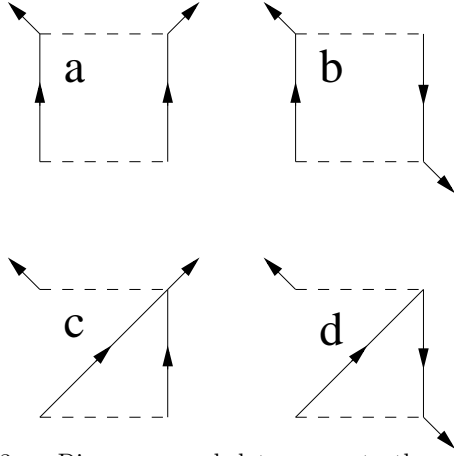} }
\caption{
{\narrowtext Diagrams needed to compute the second order
corrections at $T=0$, and at any $T$ in Wilson's scheme.
Each of the two external lines (corresponding to a $\hat{u}$ field) is
connected to a {\it tree} of response functions (the lines) due to
causality, and provides an orientation to these lines: we drew the
arrows just for clarity.}}
\label{diagsecord}
\end{figure}

The fast-modes average $\left<S_{\rm
i}^2\right>_c$ can be decomposed into one term with $\tilde{f}$ in
factor plus the rest which does not contain $\tilde{f}$. The former
vanishes for the following reason: the contraction of
the $\tilde{f}\int_{rt}i\hat {u}_{rt}$ with the $u_{r' t_{1}}$ or $u_{r'
t_{2}}$ contained in the vertex operator involves a fast response
$R^>_{r_{i}-r,t}$. But $\int_{r} R^>_{rt}=0$, since its
modes live in the shell. The latter is
the connected average of two disorder vertices.
We now extract from it a correction
to the disorder, i.e., a term which has the form $-\frac{1}{2}
\int_{rtt'}i\hat{u}_{rt}i\hat{u}_{rt'}\delta\Delta(u_{rt}-u_{rt'}+v(t-t'))$.
The corresponding diagrams are represented in Figure~\ref{diagsecord}.
Each diagram has two external $i\hat {u}_{rt}$ $i\hat {u}_{r't'}$
legs, to which corresponds a functional half vertex of $u_{rt}$ and
$u_{r't'}$ respectively. Calling $\tau $, $\tau' $ the (positive) time
arguments of both response functions, denoting
$U=u_{rt}-u_{r't'}+v (t-t')$, the diagrams have the
following analytical expressions,
integrated over $r,r',t,t',\rho,\tau,\tau '$:
\begin{eqnarray*}
a&=&-i\hat {u}_{rt}i\hat {u}_{r't'}\delta_{r'-r'}
\Delta''(U)\Delta(U+v (\tau'-\tau ))R^>_{\rho
\tau }R^>_{\rho \tau' }\nonumber \\
b&=&-i\hat {u}_{rt}i\hat {u}_{r't'}\delta_{r'-r-\rho}
\Delta'(U+v\tau' )\Delta'(U-v\tau )R^>_{\rho
\tau }R^>_{-\rho \tau' }\nonumber \\
c&=&i\hat {u}_{rt}i\hat {u}_{r't'}\delta_{r'-r}
\Delta''(U)\Delta(v (\tau'-\tau )) R^>_{\rho
\tau }R^>_{\rho \tau' }\nonumber \\
d&=&i\hat {u}_{rt}i\hat {u}_{r't'}\delta_{r'-r-\rho}
\Delta'(U+v\tau' )\Delta'(-v
(\tau'+\tau ))R^>_{\rho \tau }R^>_{-\rho \tau' }\nonumber
\end{eqnarray*}
After another short distance expansion of $b$ and $d$, noting that
$\int_{\rho }R^>_{\pm \rho \tau }R^>_{\rho \tau' }=\int_{q}R^>_{\mp q \tau
}R^>_{q \tau' }=S_{D}\Lambda ^{D}e^{- c\Lambda ^{2} (\tau '+\tau )/\eta
}dl \,/\eta ^{2}$, a proper symmetry counting yields
the term of order $\Delta ^{2}$ of (\ref{flow}).

The results obtained here are consistent with the analysis of 
Ref.~\onlinecite{moving_bragg_glass}.

\section{New results in the non-driven case at finite temperature}
\label{app:temperature}

We give here a detailed analysis of the functional
renormalization group flow at $T>0$ and zero velocity. The
temperature is an irrelevant operator and flows exponentially fast
to zero. We show however that the temperature 
rounds the cusp in a region of size proportional to
$T$ around the origin and that in this boundary layer, the disorder
correlator takes a {\it super-universal} (to lowest order in $\epsilon $)
scaling form. In addition we show how to carry a systematic expansion
at low $T$. As temperature decreases,
the correlator of the disorder becomes more and more
pinched, and eventually reaches its zero-temperature cuspy
fixed point at infinity.

We show that during the renormalization at $v=0$
with a flowing temperature $T_{l}\rightarrow 0$, the cusp forms
only asymptotically ($l\rightarrow \infty $), and $\Delta (u)$ has the
following scaling form in the
boundary layer $|u|\sim T_{l}/\chi $
\begin{eqnarray}\label{scalingform}
\Delta_{l} (u)\simeq \Delta_{l} (0)-T_{l}f (u \chi /T_{l})
\end{eqnarray}
with $f (x)=\sqrt{1+x^{2}}-1 $ and where $\chi = |\Delta^{*\prime}(0^{+})|$
measures the cusp.

Furthermore, we show that the following expansion in temperature for
the solution of the FRG flow holds
\begin{eqnarray}
(\Delta (u)-\Delta (0)-T)^{2}=\sum_{n\geq 2}T^{n}f_{n} (u/T)
\end{eqnarray}
thus we obtained a fairly complete picture of the solution.

\subsection{The curvature}
\label{exfp}

The flow equation of the value at zero of the disorder correlator is
\begin{eqnarray}\label{flowzero}
\partial_{l} \Delta_{l} (0)= (\epsilon -2\zeta )\Delta_{l} (0)
+T_{l}\Delta''_{l} (0)
\end{eqnarray}
Since $\Delta_{l}\rightarrow \Delta ^{*} $, the convergence of
$\Delta_{l} (0)$ towards $\Delta ^{*} (0)$ implies that
$T_{l}\Delta''_{l} (0)$ also converges. From the fixed point equation
\begin{eqnarray}
(\epsilon-3\zeta ) \Delta^{*} (u)
+ \zeta \left( u \Delta^{*} (u)\right)'=
\frac{1}{2}\left(\Delta^{*} (u) -\Delta^{*} (0)\right)^{2\prime \prime
}\nonumber
\end{eqnarray}
one has simply $(\epsilon -2\zeta )\Delta ^{*} (0)=\Delta
^{*\prime} (0^{+})^{2}$, and thus,
\begin{eqnarray}\label{alpha=1}
-T_{l}\Delta''_{l} (0)\rightarrow \Delta
^{*\prime} (0^{+})^{2}
\end{eqnarray}

\subsection{Scaling function in the boundary layer}
\label{scalingboundary}

We show here that the assumption that the curvature at zero of $\Delta
_{l}$ diverges like a power of the inverse temperature implies that
{\it all} the derivatives at zero also diverge and that there exists a
well defined and particularly simple scaling function in the boundary
layer around zero.

Precisely, for any function $T_{l}$ decreasing to zero and a function
$\Delta_{l}(u)$ such that
\begin{eqnarray}
\partial_{l}\Delta_{l} (u) =& (\epsilon-2\zeta ) \Delta_{l} (u) + \zeta u
\Delta'_{l} (u)+ T_{l} \Delta''_{l}(u) \\
&+ \Delta_{l}'' (u) \left(\Delta_{l} (0)-\Delta_{l} (u) \right)
-\Delta_{l}' (u)^{2} \nonumber
\end{eqnarray}
if $\Delta_{l}'' (0)\sim  -\left(\chi^{2}/T_{l} \right)^{\alpha}$
for some $\alpha >0$ and $\chi $, then, defining the functions $f_{l}
(x)=\frac{1}{T_{l}}\left( \Delta_{l} (0)-\Delta_{l}
(xT_{l}^{(\alpha +1)/2}\chi^{-\alpha })\right)$,
we obtain that every derivatives of $f_{l}$ at
$x=0$ converge to the corresponding derivatives of
$f(x)=\sqrt{1+x^{2}}-1$, and that $f$ is
the only fixed possible fixed point for $f_{l}$.

A simple way to see the convergence to the scaling function
$f$ is to write the flow of $f_{l}$
\begin{eqnarray}
&&T_{l}\Delta_{l} '' (0)+\frac{1}{2}\chi ^{2\alpha }T_{l}^{1-\alpha }
(1+f_{l})^{2\prime \prime}=\nonumber\\
&&T_{l}\left( \partial_{l}f_{l}-2
f_{l} +\frac{\alpha +1}{2}f'_{l}\right)\nonumber
\end{eqnarray}
and eliminate at large $l$ the rhs term which is subdominant (higher order in
$T_{l}$) for $\alpha >0$,
since $T$ has been absorbed in the variable $x$ of $f_{l}$. We have
used that $\theta =2-\epsilon +2\zeta $. Hence the
fixed point equation for $f_{l}$ is 
\begin{eqnarray}
\frac{1}{2}\left(1+f \right)^{2\prime
\prime}=1\nonumber
\end{eqnarray}
which has the solution $f (x)$ above since we know that
$f (0)=0$, $f'' (0)=1$ and $f^{(4)} (0) =-3$ is easily checked.

This is confirmed by the study of the flow equations for the
successive derivatives $a_{n}=\Delta ^{(2n)} (0)$:
\begin{eqnarray}\label{eqdesa}
\partial a_{n}=&(\epsilon+2(n-1)\zeta ) a_{n} \\
&+ T a_{n+1}
-\frac{1}{2}\sum_{k=1}^{n}\left(^{2 (n+1)}_{2k} \right)a_{k}a_{n+1-k} \nonumber
\end{eqnarray}
 From a trivial recurrence, the hypothesis $\Delta_{l}'' (0)\sim
-\left(\chi^{2}/T_{l} \right)^{\alpha}$ implies that $T^{n(\alpha +1)-1}a_{n}$
converges for any $n$. Moreover the limit $c_{n}=\lim_{l\rightarrow
\infty }{T^{n(\alpha +1)-1}\chi ^{-2n\alpha }a_{n}}$ can be obtained from
(\ref{eqdesa}) and is $c_{n}=\frac{(1.3\dots
(2n-1))^{2}}{2n-1}=f^{(2n)} (0)$.

To fix the value of $\alpha $ ($\alpha =1$ as strongly suggested by
(\ref{alpha=1})), we checked that the only values of
$\beta>0$, $\gamma>0$ such that
$g_{l} (x)=\frac{1}{T^\gamma_{l }}(\Delta^*(T_{l }^\beta x)
-\Delta_{l}(T_{l }^\beta x))$
has a meaningful fixed point are $(\beta,\gamma)=(1,1)$. For these
values, the fixed point is $g(x)=
\Delta^{*\prime}(0^+)x+\sqrt{1+\left( \Delta^{*\prime}(0^+)x\right)^{2}}$.

\subsection{Next order in $T$}
\label{nextt}

The procedure which gives us the leading behavior in the boundary
layer controlled by temperature can be extended analytically with
arbitrary accuracy in an expansion to any order in $T$. We study
\begin{eqnarray}
\partial_{l}\Delta_{l} (u)&=& (\epsilon-2\zeta)\Delta_{l} (u)+\zeta u
\Delta_{l}' (u) \nonumber \\
&&+ T_{l}\Delta_{l}'' (u)-\Delta_{l}'' (u)
\left(\Delta_{l} (u)\Delta_{l} (0) \right)-\Delta_{l}' (u) ^{2}\nonumber \\
\partial_{l}\ln T_{l}&=&-\theta \nonumber
\end{eqnarray}
with $\theta=2-\epsilon+2\zeta$.
For numerical purposes or for the following analytical computation,
it is useful to switch to the function $y (u)= (\Delta (u)-\Delta (0)-T)^{2}$
which remains quadratic at the origin when $T\rightarrow 0$, since $y
(u)=T^{2}+ |T\Delta'' (0)|u^{2} + {\cal O} (u^{4})$ for $T>0$ and $y (u)=
\Delta' (0^{+})^{2}u^{2} + {\cal O} (u^{4})$ for $T=0$. This function
flows as
\begin{eqnarray}
\partial_{l}y &=&2 (\epsilon-2\zeta)y+\zeta u y'
+\sqrt{y}\left(y''-y'' (0)-4T\right)\nonumber
\end{eqnarray}
We can replace the scale $l$ dependence of $y_{l} (u)$ by a $T$ dependence
since $T$ and $l $ are linked by
$T_{l}=T_{0}e^{-\theta l}$. The function $y_{T}
(u)$ can be expanded in
\begin{eqnarray}
y_{T} (u)= \sum_{n\geq 2}T^{n}f_{n} (\frac{u}{T})\nonumber
\end{eqnarray}
The expansion begins at $n=2$ since $y_{T} (0)\equiv T^{2}$ and we have
\begin{eqnarray}
f_{2} (0)=1\qquad f_{n>2} (0)=0\nonumber
\end{eqnarray}
The equation for the $f_{n}$'s reads
\begin{eqnarray}
\sum_{n\geq 2}T^{n}\left(\left(2 (\epsilon -2\zeta) +n \theta
\right)f_{n}+ (\zeta-\theta)xf_{n}' \right)= \nonumber \\
\sqrt{\sum_{n\geq 2}T^{n}f_{n}}\left( 4T-\sum_{n\geq 2}T^{n-2}
(f_{n}''-f_{n}'' (0)) \right)\nonumber
\end{eqnarray}
One can solve this equation order by order in $T$. It is useful to
divide $\Delta$ and $T$ by $\chi^{2}$ and $u$ by $\chi$. With these
rescaled quantities, we have simply
\begin{eqnarray}
y_{T_{l}}'' (0) = -2T_{l}\Delta_{l}'' (0) \rightarrow 2\nonumber
\end{eqnarray}
and thus $f_{2}'' (0)=2$. If we knew the full behavior of
$y_{T}'' (0)$, i.e., the $f_{n}'' (0)$'s, we could completely solve
the system. Here, we get
\begin{eqnarray}
f_{2} (x)&=&1+x^{2}\nonumber \\
f_{3} (x)&=&4\left(1-\frac{\epsilon-\zeta}{3}
\right)\left(\sqrt{1+x^{2}}-1-\frac{x^{2}}{2} \right)\nonumber \\
&&- (4- (\epsilon-\zeta))x (\asinh x-x)\nonumber \\
&&-\frac{\epsilon-\zeta}{3}x^{2}\sqrt{1+x^{2}}\nonumber \\
&&+f_{3}'' (0)\frac{x^{2}}{2}\nonumber
\end{eqnarray}
where we wrote $f_{3} (x)$ such that the three first lines are
functions which vanish and have zero curvature at zero. Note that
while $f_{2} (x)$ is universal, the
last term $f_{3} (x)$ contains un unknown integration constant
$f_{3}'' (0)$ which
presumably depends on the initial condition of the flow and is thus
not universal. Indeed we observed a non-universal $f_{3}'' (0)$ in a
numerical integration of the flow of $y_{T} (u)$.

The procedure can be carried to any order in $T$ and the all the $f_{n}$'s are
accessible. The unknown coefficients of the expansion
\begin{eqnarray}
-2T_{l}\Delta_{l}'' (0) =2 + \sum_{n>0}T^{n}f_{n+2}'' (0)\nonumber
\end{eqnarray}
are similarly non-universal.

Both Subsections~\ref{exfp} and \ref{scalingboundary} thus provide a
rather convincing and consistent picture for the solution of the $T>0$,
$v=0$ FRG equations (awaiting a mathematical proof).

\section{Analytical solutions at fixed temperature}
\label{app:ftfp}

We present here the analytical solutions of the fixed point equations
for RF and RP at fixed $T$. Thanks to the exact expression of 
these fixed points, we are able
to check the scaling form derived in Appendix~\ref{app:temperature}
within an ``adiabatic'' hypothesis where the running correlator at
$l$ is identified with the fixed point at $T=T_{l}$. Our families of
fixed temperature fixed points (FTFP) give back the known fixed
points at $T=0$ in both the RF\cite{fisher_functional_rg} and the
RP\cite{bragg_glass_global} cases. 
However, even if we obtain the same {\it form} (\ref{x2/2=y-1-lny}) for the
RF $T=0$ fixed point as in Ref.~\onlinecite{fisher_functional_rg}, we
disagree with the scaling in $\epsilon$.

\subsection{Random field}
\label{rft}

We look for a fixed point of
\begin{eqnarray}\label{eqrfft}
\partial_{l}\Delta_{l} (u) =& (\epsilon-3\zeta ) \Delta_{l} (u)
+ \zeta \left( u \Delta_{l} (u)\right)'\\
&-\frac{1}{2}\left(\Delta_{l} (u) -\Delta_{l} (0) -T\right)^{2\prime \prime }
\end{eqnarray}
with {\it fixed} $T$ and initial random field condition $\int \Delta
_{0}>0$. Since $\partial_{l}\ln \int du\,\Delta _{l} (u)=
\epsilon-3\zeta$, a meaningful fixed point can be obtained only
for $\zeta =\epsilon /3$. Fixing the RF strength $\int \Delta
_{0}$ to one, we are led to the following problem: for any $T\geq 0$,
find the fixed temperature fixed point function (FTFP)
$\Delta (T,u)$ such that
\begin{eqnarray}
\frac{\epsilon }{3}\left(u\Delta (u) \right)'&=&\frac{1}{2}\left( \Delta
(u) -\Delta (0) -T\right)^{2\prime \prime}\label{rf1}\\
\int du\,\Delta (u)&=&1\label{rf2}
\end{eqnarray}
Integrating (\ref{rf1}) from $0$ to $\infty $ yields $T\Delta '
(0^{+})=0$, hence the FTFP has a cusp for $T=0$ and no cusp for
$T\neq 0$.

At $T=0$, integrating (\ref{rf1}) from $0$ to $u$ and dividing by $\Delta
(u)$ yields $u=\Delta ' (u)-\Delta (0)\Delta ' (u)/\Delta (u)$. Then,
integrating again from $0$ to $u$  yields the $T=0$ FTFP, by
imposing (\ref{rf2})
\begin{eqnarray}\label{del0u}
\Delta (T=0,u)=\left(\frac{\epsilon}{3 (\int y)^{2}} \right)^{1/3} \,
y \left(u  \left(\frac{\epsilon \int y}{3} \right)^{1/3} \right)
\end{eqnarray}
where the function $y(x)$ is implicitely defined
by\cite{fisher_functional_rg}
\begin{eqnarray}\label{x2/2=y-1-lny}
\frac{x^{2}}{2}=y-1-\ln y
\end{eqnarray}
Since $y (0)=1$ one has
\begin{eqnarray}\label{del00}
\Delta (0,0)=\left(\frac{\epsilon}{3
(2\sqrt{2}\int_{0}^{1}\sqrt{y-1-\ln y})^{2}} \right)^{1/3}
\end{eqnarray}
It is easy to compute the number $\int y=\int_{-\infty }^{\infty }
dx\, y (x)=2\int_{0}^{1} dy\, x
(y)=2\sqrt{2}\int_{0}^{1}dy\,\sqrt{y-1-\ln y }\simeq 1.55$. Note the
behavior near $0$ given by $y
(x)=1-|x|+\frac{x^{2}}{3}-\frac{|x^{3}|}{36}+ {\cal O} (x^{4})$
thus $\partial_{u}^{2}\Delta (0,0^{+})=\frac{2\epsilon}{9}$.
Note also the gaussian
decrease of correlations at infinity $y (x)\sim e ^{ -1-x^{2}}$.

An intriguing fact is the scaling of the $T=0 $ fixed point with
$\epsilon$: its $n^{\rm th}$ derivative at $0^{+}$ scales like
\begin{eqnarray}
\partial_{u}^{n}\Delta (T=0,u=0^{+})\sim \epsilon ^{ (1+n)/3}
\end{eqnarray}

\begin{figure}[htb]
\centerline{\fig{8cm}{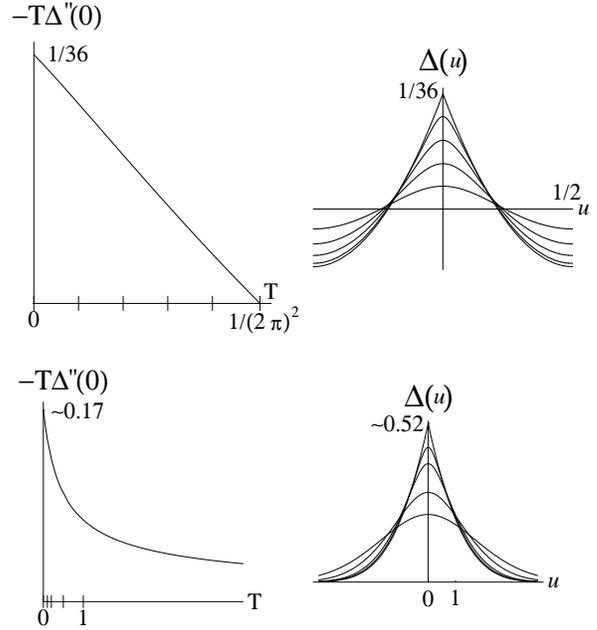}}
\caption{{\narrowtext Fixed points at fixed temperature $\Delta
(T,u)$. Bottom: RF case, the solution $\Delta
(T,u)$ to the reduced equation (\ref{eq:rfred1},\ref{eq:rfred2})
exists for any $T$. Right bottom: plot of the RF FTFPs 
for $T\in \{ 0, 0.1, 0.2, 0.5, 1\}$, these temperature are located on
the abscissas of the left bottom plot.
On the left, plot of $-T\partial^{2}_{u}\Delta
(T,u=0)$ versus $T$. Top: RP case, the $1$-periodic
non-trivial solution $\Delta
(T,u)$ to (\ref{eq:rpred}) exists for
$0\leq T< (2\pi )^{-2}$. Right bottom: plot of the RP FTFPs 
for $T\in \{ 0, 0.005, 0.01, 0.015, 0.02\}$, these temperature are located on
the abscissas of the left top plot. The FTFPs are
analytic for $T>0$ but tend to the cuspy $\Delta
(T=0,u)$ as
$T\rightarrow 0$. The curve on the left shows
$-T\partial^{2}_{u}\Delta
(T,u=0)$ as a function of $T$. It is {\it
not} a straight line}.}
\label{fig:fixedt}
\end{figure}

At $T>0$, there is no cusp ($\partial_{u}\Delta (T>0,0^{+})=0$) and the same
double integration of (\ref{rf1}) yields
\begin{eqnarray}
\Delta (T,u)=\Delta (T,0) y (T,u\sqrt{\epsilon / (3\Delta (0))})
\end{eqnarray}
with $y (T,x)$ implicitely defined by
\begin{eqnarray}\label{ytx}
\frac{x^{2}}{2}=y -1- (1+\frac{T}{\Delta (0)})\ln y
\end{eqnarray}
The value of $\Delta (T,0)$ is determined by condition (\ref{rf2}).
Using $\int dx \, y (x)=\int dy \, x
(y)$, this condition reads
\begin{eqnarray}\label{defdel0}
\sqrt{\frac{24 \Delta (T,0)^{3}}{\epsilon}}\int_{0}^{1}\!\!\!\!\!dy\,\sqrt{
y-1- (1+\frac{T}{\Delta (T,0)})\ln y} =1
\end{eqnarray}
This equation admits a unique solution $\Delta (T,0)>0$ for any $T> 0$.
Then there exists a unique FTFP $\Delta (T,u)$ for each $T>0$.
Some of them are displayed
in Figure~\ref{fig:fixedt}. Note that $T=0$ in
(\ref{ytx},\ref{defdel0}) gives back the
$T=0$ non-analytic fixed point $\Delta (T=0,u)$ (\ref{del0u}). Hence
the set of FTFP has a nice $T\rightarrow 0$ limit, even if
there is a qualitative difference between the cuspy $T=0$ FTFP and the
analytic $T>0$ FTFPs.

As is obvious from their analytical expression, or from
Figure~\ref{fig:fixedt}, $\lim_{T\rightarrow 0}\Delta (T,u)=\Delta
(0,u)$:
the $T=0$ non-analytic fixed point is approached
smoothly by the set of analytic fixed $T$ fixed points.
When $T$ approaches zero, the curvature of the FTFPs at
the origin goes to $-\infty $ like
\begin{eqnarray}
\lim_{T\rightarrow 0}{-T\Delta '' (T,0)}=\frac{\epsilon }{3}\Delta (0,0)
\end{eqnarray}
with $\Delta (0,0)$ given by (\ref{del00}).

We also checked that the $\Delta (T,u)$
converge when $T\rightarrow 0$
to the zero temperature fixed point with the predicted scaling form
(\ref{scalingform})
\begin{eqnarray}
\frac{\Delta (T,0)-\Delta (T,T\frac{x}{\chi })}{T}
\stackrel{T\rightarrow 0}{\rightarrow}
\sqrt{1+x^{2}}-1
\end{eqnarray}
where $\chi = |\partial_{u}\Delta (0,0^{+})| $ is given by the $T=0$
FTFP equation $\chi^{2}=\frac{\epsilon }{3} \Delta (0,0)$.

Some of the RF fixed $T $ fixed points are shown on the right bottom
quarter of Figure~\ref{fig:fixedt}, including the cuspy (highest) $T=0$
fixed point. Absorbing $\epsilon$ in $T$ and $\Delta $,
we chose to plot the non-trivial
solution to the most reduced problem
\begin{eqnarray}\label{eq:rfred1}
\frac{1}{3} \left( u\Delta (u)\right)^{\prime}&=&\frac{1}{2}\left(\Delta
(u)-\Delta (0)-T \right)^{2\prime \prime}\\
\int du\,\Delta (u)&=&1\label{eq:rfred2}
\end{eqnarray}
To restore $\epsilon $ and $\int \Delta $, one simply has to note that
the ``dimensions'' are $T\equiv \Delta \equiv \epsilon
^{1/3}\left(\int \Delta \right)^{2/3}$ and $u\equiv \epsilon
^{-1/3}\left(\int \Delta  \right)^{1/3}$.
The left bottom of Figure~\ref{fig:fixedt} shows $-T\Delta'' (T,0)$ as a
function of $T$. This combination has a finite limit ($\simeq 0.17$) when
$T\rightarrow 0$.

\subsection{Random periodic}
\label{rfp}

In the Random Periodic case, the conservation of the period $a$ of
$\Delta $ requires $\zeta =0$. After a suitable rescaling,
$u\rightarrow u/a$, $\Delta \rightarrow \Delta / (\epsilon a^{2})$ and
$T\rightarrow T/(\epsilon a^{2})$, the fixed point equation reads for
the $1$-periodic function $\Delta (u)$
\begin{eqnarray}\label{eq:rpred}
\Delta (u)=
\frac{1}{2}\left(\Delta (u) - \Delta (0) - T \right)^{2\prime \prime }
\end{eqnarray}
and is easily solved by quadrature, by analogy with a particle's
position $X (u)=\left(\Delta (u) -\Delta (0) -T\right)^2$
at time $u$ in a potential $V (X)=4X^{3/2}/3-2 (\Delta
(0)+T)X$ verifying $X'' (u)=-V' (X (u))$.
The quadrature leads to the reciprocical function $u (X)$,
parametrized by $\Delta (0)$ and $T$, as a sum of two elliptic functions.
Then, imposing the solution $\Delta (u)$ be $1$-periodic
fixes $\Delta (0)$ as a function of $T$.

The result is:
\begin{itemize}
\item for $T\geq (2\pi)^{-2}$ the only
solution is $\Delta (u)\equiv 0$
\item for $0<T< (2\pi)^{-2}$ another
solution arises, which resembles a cosinus function of linearly
vanishing amplitude when $T\rightarrow (2\pi)^{-2}$. This non-trivial
solution has no cusp but becomes pinched as $T$ decreases
(growing curvature $|\Delta '' (0)|$ and higher harmonics).
As can be seen on the analytical
expression (not given here), $-T\Delta''
(0)\stackrel{T\rightarrow 0}{\longrightarrow}1/36$.
In particular, it remains finite when the
temperature vanishes.
\item eventually for $T\rightarrow 0$, the non-trivial solution 
uniformly tends to the zero temperature fixed point
\begin{eqnarray}
\Delta (u)=\frac{1}{6} (\frac{1}{6}-u(1-u))
\end{eqnarray}
\end{itemize}

The temperature $(2\pi)^{-2}$ in our units is {\it exactly} the
critical temperature $T_{g}$ of the random-field 
XY-model\cite{cardy_desordre_rg}
and the fixed points near $T_{g}^{-}$ reproduce the line of fixed
points of this problem (since we worked to second order, it is only an
approximation). Indeed in $D=2$, the naive dimension of the
temperature is zero and our FTFP has a direct physical
meaning. Note that another random gradient term becomes relevant in $D=2$ but
does not feed back on the flow of $\Delta
(u)$\cite{cardy_desordre_rg,carpentier_ledou_triangco}.

We can now use the exact FTFPs to check that an adiabatic hypothesis 
is consistent with the scaling form
(\ref{scalingform}). Indeed, one can numerically 
check that the correlator with a flowing
temperature has the FTFPs have the scaling (\ref{scalingform}) as
$T\rightarrow 0$.
To conclude about the problem with a flowing temperature
$T_{l}\rightarrow 0$, it appears from these observations that
no cusp occurs at finite scale for $T_{0}>0$. The cusp forms
only asymptotically ($l\rightarrow \infty $), with
\begin{eqnarray}
\lim_{l\rightarrow \infty }{-T_{l}\Delta_{l} '' (0)}=\Delta (0,0)\nonumber
\end{eqnarray}
given by (\ref{del00}) and it obeys a scaling form in the
boundary layer $|u|<T_{l}/\chi $
\begin{eqnarray}\label{scform}
\Delta_{l} (u)\simeq \Delta_{l} (0)-T_{l}f (u \chi /T_{l}),\,\,\,\,\,
f (x)=\sqrt{1+x^{2}}-1 \nonumber
\end{eqnarray}
We note that the precise form of the flow of the temperature (i.e. the
value of $\theta $) only affects subdominant behavior (i.e. the
function $f_{3} (x)$ in Subsection~(\ref{nextt})).

\section{Multi-dimensional case}
\label{app:n>1}

We give here a possible extension of the FRG to a multi-dimensional
displacement field. This study
generalizes the approach of Ref.~\onlinecite{ertas_kardar_anisotropic} by
including the effect of $v>0$ and $T>0$ 
in the flow. For periodic structures, a similar 
study of the multi-dimensional displacement field was shown in
Ref.\onlinecite{moving_bragg_glass} to yield novel effects.

In a $D+N$ dimensional space, we distinguish between the {\it
internal} or {\it longitudinal} space of
dimension $D$, to which $r$ belongs,
and the {\it transverse} space of dimension $N$, to which $u$
belongs. The elastic energy of an interface without overhangs defined
by a height function $u_{r}$ is quadratic in $\nabla u$ of the form
$\frac{1}{2}\int_{r} c_{ij}^{\ \ \mu \nu} (\partial_{\mu}u^{i}_{r})
(\partial_{\nu}u^{j}_{r})$.

The disorder: the random bond (RB) case corresponds to a
random potential $V (r,u)$, with correlations $\overline{(V (r,u)-V
(r',u'))^{2}}=-2\delta^{D} (r-r') {\sf R} (u-u')$. Function ${\sf R}
(u)$ is even, vanishes at $u=0$ and goes to a negative constant for $|u|\gg
r_{f}$. The
random field (RF) case corresponds to a force $F^{i} (r,u)$ with
correlations $\overline{F^{i} (r,u)F^{j} (r',u')}=\delta^{D} (r-r')
\Delta^{ij} (u-u')$, where the $\Delta^{ij} (u)$ vanish for $|u|\gg r_{f}$.
A RB gives rise to a RF via $F^{i}=-\partial^{i}V$ and the
correlators are related by $\Delta^{ij} (u)=-\partial^{ij}{\sf R} (u)$. Note
that this type of correlator deriving from a RB has $\int
d^{N}u\, \Delta^{ij} (u)=0$.
Finally, the random periodic case (RP) occurs
when $u$ is defined up to a discrete set of translations
forming a lattice of points $P$, e.g.
when $u$ is a phase, defined up to $2\pi$ shifts. In this case, the
disorder is periodic and one has $\Delta^{ij} (u)= \Delta^{ij} (u+P)$
for any $P$ of the lattice (or ${\sf R}(u)= {\sf R}(u+P)$).

The overdamped dynamics is given by
\begin{eqnarray}\label{eq:d+n}
\eta^{i}_{\ j}\partial_{t}u^{j}_{rt}=c^{i\ \ \mu \nu}_{\ j}
\partial_{\mu \nu}u^{j}_{rt}+F^{i}
(r,u_{rt})+\zeta^{i}_{rt}+f^{i}+h^{i}_{rt}\nonumber 
\end{eqnarray}
where $\eta$ is the friction tensor and $\zeta$ a Langevin noise,
with correlations $\langle \zeta^{i}_{rt}\zeta^{j}_{r't'}\rangle =2
(\eta T)^{ij}\delta (r-r')\delta (t-t')$. The tensor $T$ stands for
the temperature(s) of this out of equilibrium system. We added a
driving force $f^{ i}$ perpendicular to the interface and a source
field $h^{i }_{rt}$, as an external excitation.

Without assuming any symmetry, let $C^{ij}_{rt}$ and $R^{ij}_{rt}$
be the gaussian correlation and response functions. We obtain by the
same procedure as for the $N=1$ case the following first order
corrections due to disorder
\begin{eqnarray}
\delta c^{ij}_{\mu \nu }&=&0\nonumber \\
\delta \tilde{f}^{i}&=&\int_{\kappa t}e^{i\kappa .(C_{00}-C_{0t}).i\kappa
+i\kappa .v t}\Delta _{\kappa}^{ik
}i\kappa^l R_t^{l k}\nonumber \\
\delta \eta ^{ij}&=&-\int_{\kappa t}e^{i\kappa .(C_{00}-C_{0t}) . i\kappa
+i\kappa .v t}\Delta _{\kappa}^{ik}i\kappa^l t R_t^{lk} i\kappa^j\nonumber \\
\delta (\eta T)^{ij}&=&\frac{1}{2} \int_{\kappa t}e^{i\kappa .vt}
(e^{i\kappa.(C_{00}-C_{0t}). i\kappa}-e^{i\kappa .C_{00}. i\kappa})
\Delta ^{ij}_\kappa\nonumber \\
\delta \Delta ^{ij}(u)&=&\int_\kappa \Delta ^{ij}_\kappa
e^{i\kappa .C_{00}. i\kappa +i\kappa .u}\nonumber
\end{eqnarray}

Using $(\Delta_{\kappa}^{\alpha \beta})^{*}=\Delta_{-\kappa}^{\alpha
\beta}$, $\Delta_{-\kappa}^{\alpha \beta}=\Delta_{\kappa}^{
\beta \alpha}$, we write the on-shell corrections as (with the matrix
product $A.B=A_{\alpha \gamma }B_{\gamma \beta }$)
\begin{eqnarray}
\delta c^{ij}_{\mu \nu }&=&0\nonumber \\
\delta \tilde{f}&=&-\int_{\kappa} i\kappa.\int_{t} R^{>}_{0t}.\Delta
_{-\kappa}e^{i\kappa .vt}\nonumber \\
\delta \eta &=&-\int_{\kappa}
i\kappa.\int_{t}tR^{>}_{0t}.\Delta _{-\kappa}e^{i\kappa .vt}i\kappa
\nonumber \\
\delta (2\eta. T)&=&-\int_{\kappa}
i\kappa.\int_{t}C^{>}_{0t}.i\kappa \Delta _{\kappa}e^{i\kappa .vt}\nonumber \\
\delta \Delta_\kappa&=&i\kappa.C^{>}_{00}.i\kappa \Delta _{\kappa}\nonumber
\end{eqnarray}

The second order correction to $\Delta $ reads
\begin{eqnarray}
\delta \Delta ^{i j }(u)=\int_{q\tau \tau'}
R^{>m k}_{q\tau } R^{>m' l}_{q\tau'}  \nonumber \\
\left[
\left( \Delta ^{k l}(v(\tau '-\tau))-\Delta ^{k l}(u+v(\tau'-\tau )) \right)
\partial^{m }\partial^{m'}\Delta ^{i j }(u)\right.\nonumber \\
-\partial^{m }\Delta ^{i l}(u+v\tau )
 \partial^{m'}\Delta ^{k j }(u-v\tau')\nonumber \\
-\partial^{m }\Delta ^{i l}(u+v\tau )
 \partial^{m'}\Delta ^{j k }(v(\tau'+\tau))\nonumber \\
 \left.+
 \partial^{m }\Delta ^{i l}(v(\tau'+\tau))
 \partial^{m'}\Delta ^{k j }(u-v\tau') \right]\nonumber
\end{eqnarray}

Note that each of the first three terms are symmetric under $i
\leftrightarrow j ,u\leftrightarrow -u$ and that the fourth is exchanged
with the fifth under this symmetry. Then $\Delta $ remains a correlator.

Of course this second order correction to $\Delta$ gives back the expression
already computed for a $D+1$ interface if $N=1$. At zero velocity, one
gets the second derivative of the flow equation of
Balents and Fisher \cite{balents_frg_largen}.
If we assume that
$\Delta (u)$ depends only on the component of $u$ parallel to the
velocity and send $v$ to zero,
then our expression reduces to the equations of Ertas and
Kardar \cite{ertas_kardar_anisotropic}.

To simplify the analysis, let us rely on the assumed symmetries of the
system. If we suppose that the initial problem is rotationnally invariant,
i.e. has ${\rm O}(N)$ symmetry, then the elasticity tensor $c$, the friction
tensor $\eta $ and the temperature tensor $T$ are only scalars and the
force--force correlator $\Delta $ is covariant, {\it i.e.}, for any ${\cal R}$
such that ${\cal R}^{\dagger }.{\cal R}=1$, ${\cal R}^{\dagger}.\Delta
(u).{\cal R}=\Delta({\cal R}.u)$.

During the flow, we expect from physical grounds
that the running terms of the action will conserve their
symmetries but the velocity $v$ which
is fixed once for all selects a particular direction in
transverse space. The interesting symmetries
are given by the little group of the velocity, i.e., the
transformations ${\cal R}$ such that ${\cal R}^{\dagger }.{\cal R}=1$ and
${\cal R}.v=v$. Then one may decompose the tensors on a basis
involving $v$ (one has only two frictions, temperatures, response and
correlation functions and five\cite{boulette} $\Delta _i$'s, functions of
$(u^2,v^2,u.v)$).

Unfortunately, the full problem can not be easily decoupled, even with
the simplifications pointed out above. No closed equation e.g. for the
correlator restricted to displacements aligned with the velocity
$\Delta (u\parallel v)$, has been found, and the problem even at zero
temperature seems involved. The simplification used in 
Ref.~\onlinecite{ertas_kardar_anisotropic} 
consists in assuming that $\Delta $ does not depend on
the transverse coordinates. This assumption reduces the problem to the
$N=1$ case, and it would be interesting to solve at finite $T$ the
behavior of transverse coordinates along the lines of our analysis.

\section{The flow of the disorder correlator at small velocity}
\label{app:v}

The effect of a small velocity on the FRG flow is mainly restricted to
the boundary layer of width $\rho _{l}$ about the origin.
Analytically, it is rather difficult to give an estimate of $\rho
_{l}$ or to decide how $\tilde{\Delta} _{l} (u)$ precisely
behaves in the boundary layer $|u|\sim \rho _{l}$. It is however
possible to simplify the formidable second order correction to the
disorder correlator, displayed in (\ref{flow}), and to obtain
analytically several results, giving some hints about this
behavior. 

The ${\cal O} (\tilde{\Delta} ^{2})$ term in (\ref{flow}) is written
under a form involving two integrations over $s,s'$, reflecting the
presence of two response functions integrated over time. After some
integrations by part, the ${\cal O} (\tilde{\Delta} ^{2})$ term
becomes\cite{emig} 
\begin{eqnarray}\label{oneint}
&&\tilde{\Delta} ^{\prime\prime}(u)\int_{s>0}\!\!\!\!\!\!
e^{-s}\left( \tilde{\Delta} (\lambda s)-\frac{\tilde{\Delta}
(u+\lambda s)+\tilde{\Delta} (u-\lambda s)}{2}\right)\\
&&+\int_{s>0}\!\!\!\!\!\!
e^{-s}\frac{\tilde{\Delta} (u+\lambda s)-\tilde{\Delta} (u)}{\lambda}
\,\,\,\int_{s>0}\!\!\!\!\!\!
e^{-s}\frac{\tilde{\Delta} (u-\lambda s)-\tilde{\Delta}
(u)}{\lambda} \nonumber\\
&&-\int_{s>0}\!\!\!\!\!\!
e^{-s}\tilde{\Delta} '(\lambda s)\frac{\tilde{\Delta} (u+\lambda
s)+\tilde{\Delta}
(u-\lambda s)-2\tilde{\Delta} (u)}{\lambda}\nonumber
\end{eqnarray}
Integrated over $u$, this correction becomes
\begin{eqnarray}
\int_{0}^{\infty}du\,\tilde{\Delta}
^{\prime}(u)\int_{s>0}e^{-s}(2-s)\frac{\tilde{\Delta} (u+\lambda s)-\tilde{\Delta} (u-\lambda
s)}{\lambda}\nonumber
\end{eqnarray}
For any non-crazy function $\Delta$, this expression is
positive. Assuming that $\Delta $ has no cusp, it can be safely
expanded and we can check that it is of order $\lambda ^{2}$:
\begin{eqnarray}
\partial\int \tilde{\Delta} = (\epsilon - 3\zeta)\int \tilde{\Delta} +
2\lambda^{2}\int \tilde{\Delta}^{\prime \prime 2} + {\cal  O}
(\lambda^{4})\nonumber
\end{eqnarray}
Thus at $v\neq 0$,
the integral of $\tilde{\Delta}$ {\it grows} during the flow, whereas
it was {\it conserved} in the statics.

Using (\ref{oneint}), one can also compare the flow of
$\tilde{\Delta}_{l}$ at small velocity to the cuspy $v=0$
flow. In particular, one observes that the effect of the velocity is
to reduce the blow-up of the curvature
$\tilde{\Delta}^{\prime \prime}(0)$
\begin{eqnarray}
\partial \tilde{\Delta}^{\prime \prime}(0)&=&\epsilon
\tilde{\Delta}^{\prime \prime}(0)
-\tilde{\Delta} ^{\prime \prime}(0)\int_{s>0}\!\!\!\!\!\!e^{-s}\frac
{\tilde{\Delta} (\lambda s)-\tilde{\Delta} (0)}{\lambda^{2}}
\nonumber\\
&&
-\int_{s>0}\!\!\!\!\!\!e^{-s}
\left( \frac{\tilde{\Delta} ^{\prime}(\lambda s)}{\lambda }
\right)^{2}\nonumber\\
&=&\epsilon \tilde{\Delta}^{\prime \prime}(0)
-3 \tilde{\Delta}^{\prime \prime}(0)^{2} \nonumber\\
&&- 9\lambda^{2}
\tilde{\Delta}^{\prime \prime}(0)\tilde{\Delta}^{\rm iv}(0)+ {\cal O}
(\lambda^{4})\nonumber
\end{eqnarray}
(note that $\tilde{\Delta} '' (0)<0$ whereas
$\tilde{\Delta} ^{\rm iv}(0)>0$). The flow of the friction 
is similarly slowed down:
\begin{eqnarray}
\partial \ln \eta &=&\int_{s>0}e^{-s}(2-s)\frac{\tilde{\Delta} (\lambda s)-
\tilde{\Delta} (0)}{\lambda ^{2}}\nonumber\\
&=&-\tilde{\Delta}^{\prime \prime}(0) -3 \lambda
^{2}\tilde{\Delta}^{\rm iv}(0) +  {\cal O} (\lambda^{4})\nonumber
\end{eqnarray}

The term $-\tilde{\Delta}' (0^{+})^{2}$ 
in the $v=0$ flow of $\tilde{\Delta}(0)$ in (\ref{deltav0}) at $v=0$
is replaced at $v>0$ by
\begin{eqnarray}\label{slowdel0}
\partial \tilde{\Delta} (0)&=&(\epsilon -2\zeta )\tilde{\Delta} (0)
\\
&&+\left(\int_{s>0}\!\!\!\!\!\!
e^{-s}\frac{\tilde{\Delta} (\lambda s)}{\lambda
}\right)^{2}-\int_{s>0}\!\!\!\!\!\!
e^{-s}\left( \frac{\tilde{\Delta} (\lambda s)}{\lambda }
\right)^{2}\nonumber
\end{eqnarray}
which has the right sign (using Cauchy inequality) to slow down the
exponential growth of $\tilde{\Delta}_{l}(0)$.

Obtaining a numerical integration of the flow is a highly non-trivial
quest\cite{emig},
since all the interesting properties occur close to the
origin yielding unaccurate results in real space. In Fourier space,
the number of harmonics to be retained is huge if one wants to focus
on the quasi-cuspy behavior ($\Delta^{*}_{\kappa}\sim \kappa^{-2}$ at the
cuspy fixed point). However, we obtained, at least at the
beginning of the flow (up to $l_{c}$) with small initial velocity, the
curve shown in Figure~\ref{rbrfpic}. The initial condition was a RB
disorder (full line). It is obvious on the snapshot (dotted line)
close to $l_{c}$ that the flow transformed the RB into a RF.

\section{Before the Larkin length}
\label{app:before}

We show here that at the scale $l_{c}$, $\tilde{\Delta}_{l} (u)$ is
very close to the static zero-temperature fixed point
$\Delta^{*}(u)$. This can be checked numerically, even in the presence
of a small temperature and small velocity. Analytically, one cannot
obtain an exact integration of the flow, but we can compare
$\tilde{\Delta}_{l} (u)$ to the known $\Delta^{*}(u)$ by the following
arguments.

Let us take e.g. the RF case for which $\zeta =\epsilon /3$.
For weak disorder one obtains from the integration of (\ref{closed}):
\begin{equation}
e^{\zeta l_{c}}=\left(1+\frac{\epsilon }{3|\tilde{\Delta}_{0} '' (0)
|} \right)^{1/3}\simeq r_{f} \left(\frac{\epsilon }{\int \tilde{\Delta} }
\right)^{1/3}
\end{equation}
where we have used
$|\tilde{\Delta}_{0}''(0)|\simeq \tilde{\Delta}_{0}(0)/r_{f}^{2}\simeq 
\int \tilde{\Delta}/r_{f}^{3}$. 
We prove in Appendix~\ref{rft} that 
$\chi=|\Delta ^{*\prime} (0^{+})|$ verifies
$\chi \simeq \epsilon ^{2/3}\left(\int \tilde{\Delta} \right)^{1/3}$
and that $\Delta ^{*} (0)\approx \epsilon ^{1/3}\left(\int
\tilde{\Delta } \right)^{2/3}$. Thus the range 
$r_{f}^{*}\approx \Delta ^{*} (0)/|\Delta^{\prime *} (0^{+})|$ of 
$\Delta^{*}$ verifies
\begin{eqnarray}\label{chirfrf}
r_{f}^{*}\approx r_{f} e^{-\zeta l_{c}}\approx \chi /\epsilon
\end{eqnarray}
To determine the range of $\tilde{\Delta}_{l_c} (u)$ we use the fact
that at the beginning of the
flow one can neglect the non-linear term in the flow equation (\ref{flow}).
We are left with $\tilde{\Delta} _{l} (u)=e^{(\epsilon
-2\zeta)l}\tilde{\Delta} _{0} (ue^{\zeta
l})$ and thus the range of $\tilde{\Delta} _{l_{c}} (u)$ is simply
\begin{equation} \label{rangelc}
r_{f} (l_{c}) \simeq r_{f}e^{-\zeta l_{c}}
\end{equation}
A comparison of (\ref{chirfrf}) and (\ref{rangelc}) shows that
the two ranges are similar. Furthermore, in the RF case $\int \Delta$
is conserved by the flow at $v=0$, and thus the similarity of ranges
shows that the shape of $\tilde{\Delta} _{l_{c}}$
is close to the shape of $\Delta^{*}$ (same integral, same range).

Similarly in the RP case it is also true that $\tilde{\Delta}
_{l_{c}}$ resembles 
$\Delta^{*}$, but now $\chi = \epsilon a/6$ as can be seen on the
fixed point
$\Delta ^{*} (u) =\frac{\epsilon }{6}\left(\frac{a^{2}}{6}-u (a-u) \right)$.


\begin{thebibliography}{10}
\bibitem{arf}
Note that for periodic structures we consider in this paper for simplicity only
  the case $r_f\sim a$ and thus the intermediate random manifold regime is
  absent\cite{bragg_glass_global}. The method can be straightforwardly extended
  to the general case.
\bibitem{asymmetry}
Note that although the initial problem does not distinguish between internal
  ($\parallel r$) and height ($\parallel u$) directions, an asymmetry occurs
  during the coarse-graining and it is useful to consider (\ref{delta}) as a
  general starting point.
\bibitem{bare}
As is shown below, the distinction RF/RB defined here for the starting model is
  modified under coarse-graining.
\bibitem{noteonkpz}
The large $v$ limit is simple only for the single component model when setting
  the internal space cutoff to zero. More generally, in this far from
  equilibrium regime, other effects can become important, such as the
  generation of non-linear KPZ type terms, or for periodic structures, effects
  related to the existence of a moving glass, as discussed
  elsewhere\cite{moving_bragg_glass,kardar_parisi_zhang}.
\bibitem{etainfini}
The divergence of the friction in the dynamical study of the statics ($f=0$,
  $v=0$) does not quantitatively feedback on the static observables but
  provides the qualitative information that the system is pinned in the sense
  that an infinite time is needed to escape the metastable states.
\bibitem{r0rd}
The scale $R_d$ in the FRG, corresponding to the onset of depinning-like
  motion, is naturally identified with $R_0$, and the FRG relation $R_d\gtrsim
  R_T$ compares with the scaling relation $R_0\gtrsim R_{\rm opt}$.
\bibitem{boulette}
One has $\Delta ^{ij}(u,v) = \Delta _1 \delta^{ij} + \Delta _2
  \hat{u}^{i}\hat{u}^{j} + \Delta _3 \hat{v}^{i}\hat{v}^{j} + \Delta _4
  (\hat{u}^{i}\hat{v}^{j} + \hat{u}^{j}\hat{v}^{i}) + \Delta _5
  (\hat{u}^{i}\hat{v}^{j} - \hat{u}^{j}\hat{v}^{i})$. Note that in
  Ref.~\onlinecite{moving_bragg_glass}, the fifth term does not appear.
\end{thebibliography}

\end{multicols}
\end{document}